%% file: ch.tex
\documentstyle[12pt]{article}

\newcommand{\mysection}{\setcounter{equation}{0}\section}

\begin{document}
\include{ch0}

\include{ch1}

\include{ch2}
\include{ch3}

\appendix
\include{cha}

\include{chr}

\include{chf}

\setlength{\unitlength}{1cm}
\begin{figure}[p]
\begin{center}
\begin{picture}(15,20)
\put(-2,0){\includegraphics{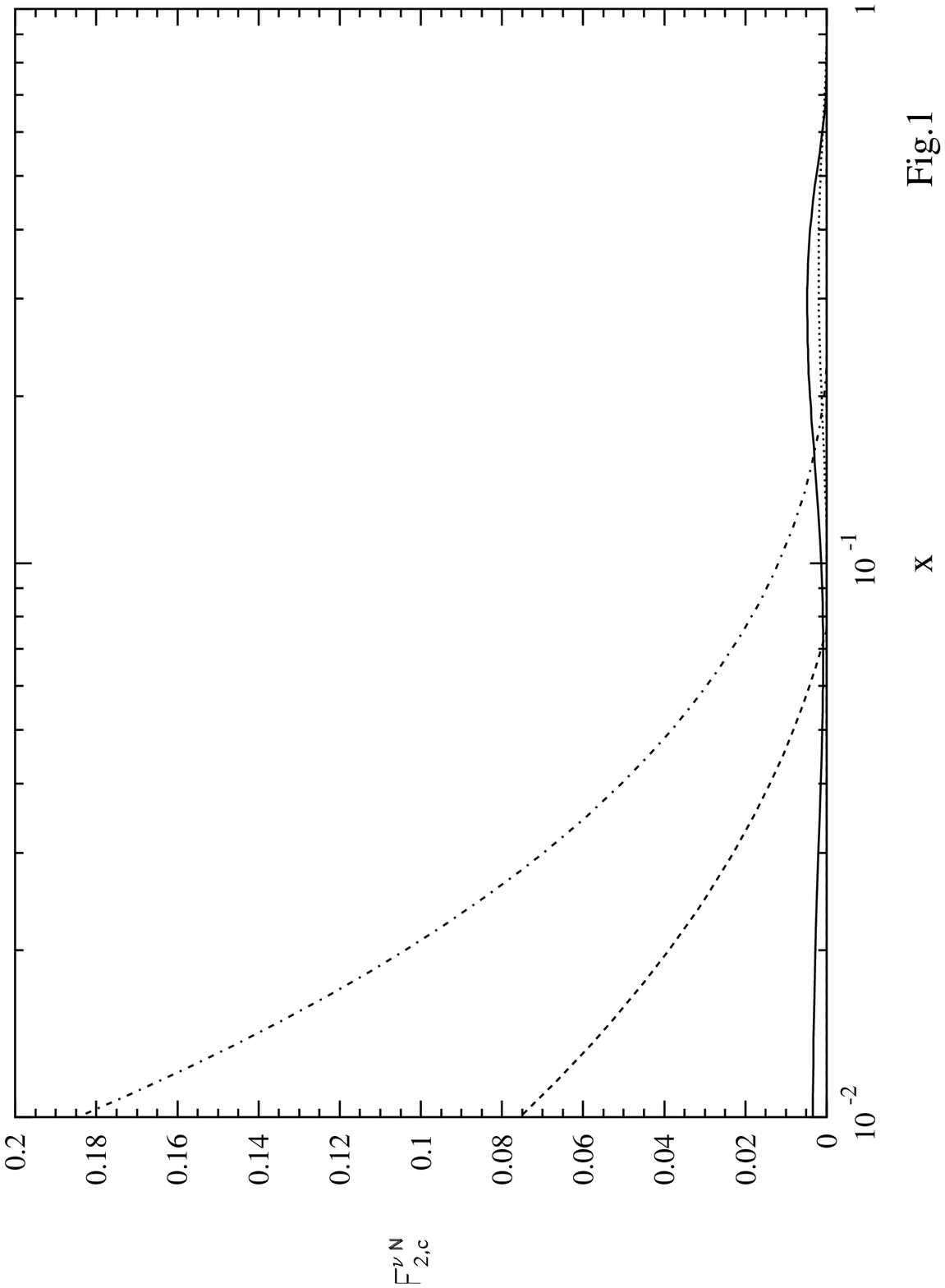}}
\end{picture}
\end{center}
\end{figure}
\begin{figure}[p]
\begin{center}
\begin{picture}(15,20)
\put(-2,0){\includegraphics{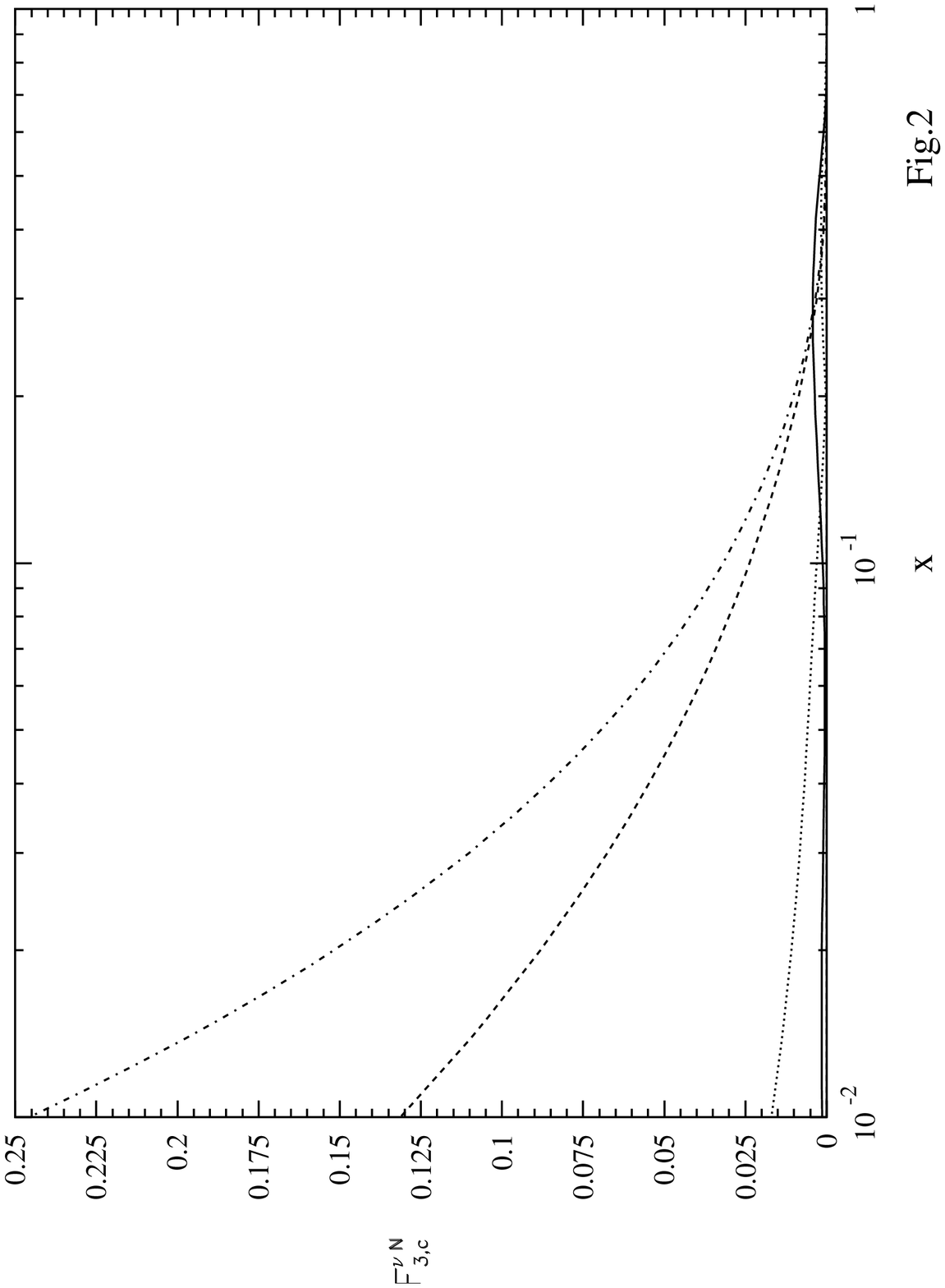}}
\end{picture}
\end{center}
\end{figure}
\begin{figure}[p]
\begin{center}
\begin{picture}(15,20)
\put(-2,0){\includegraphics{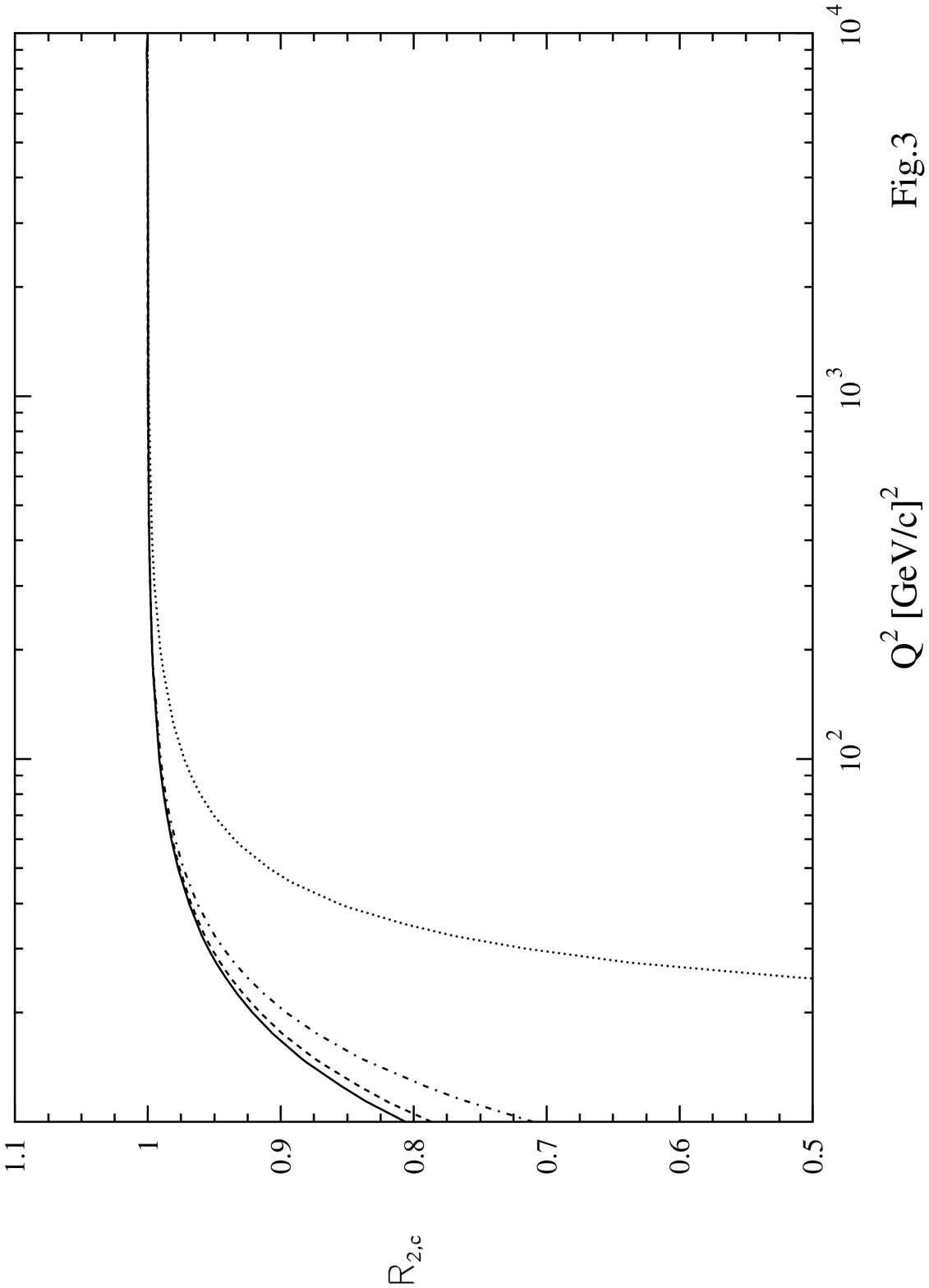}}
\end{picture}
\end{center}
\end{figure}
\begin{figure}[p]
\begin{center}
\begin{picture}(15,20)
\put(-2,0){\includegraphics{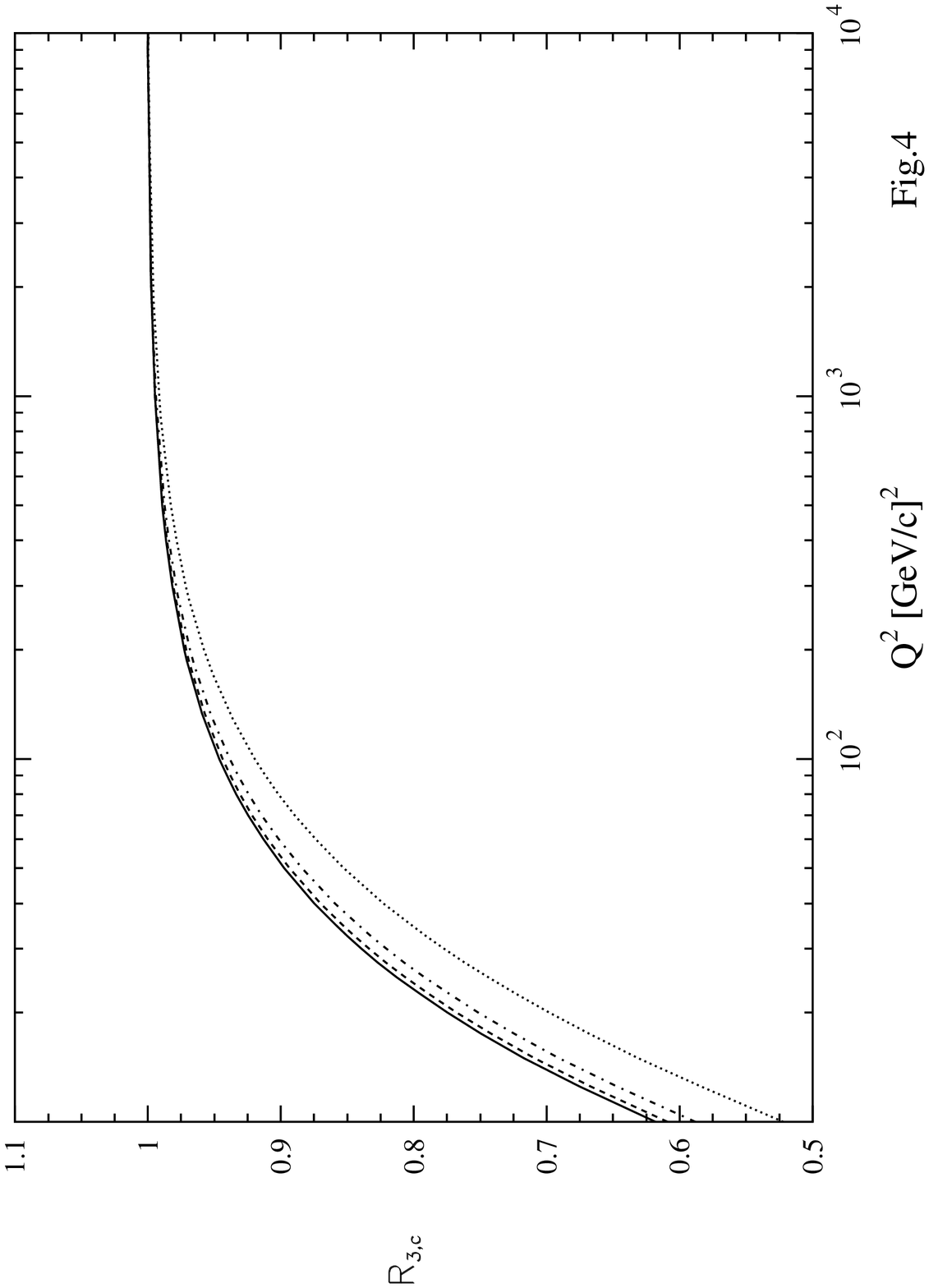}}
\end{picture}
\end{center}
\end{figure}
\begin{figure}[p]
\begin{center}
\begin{picture}(15,20)
\put(-2,0){\includegraphics{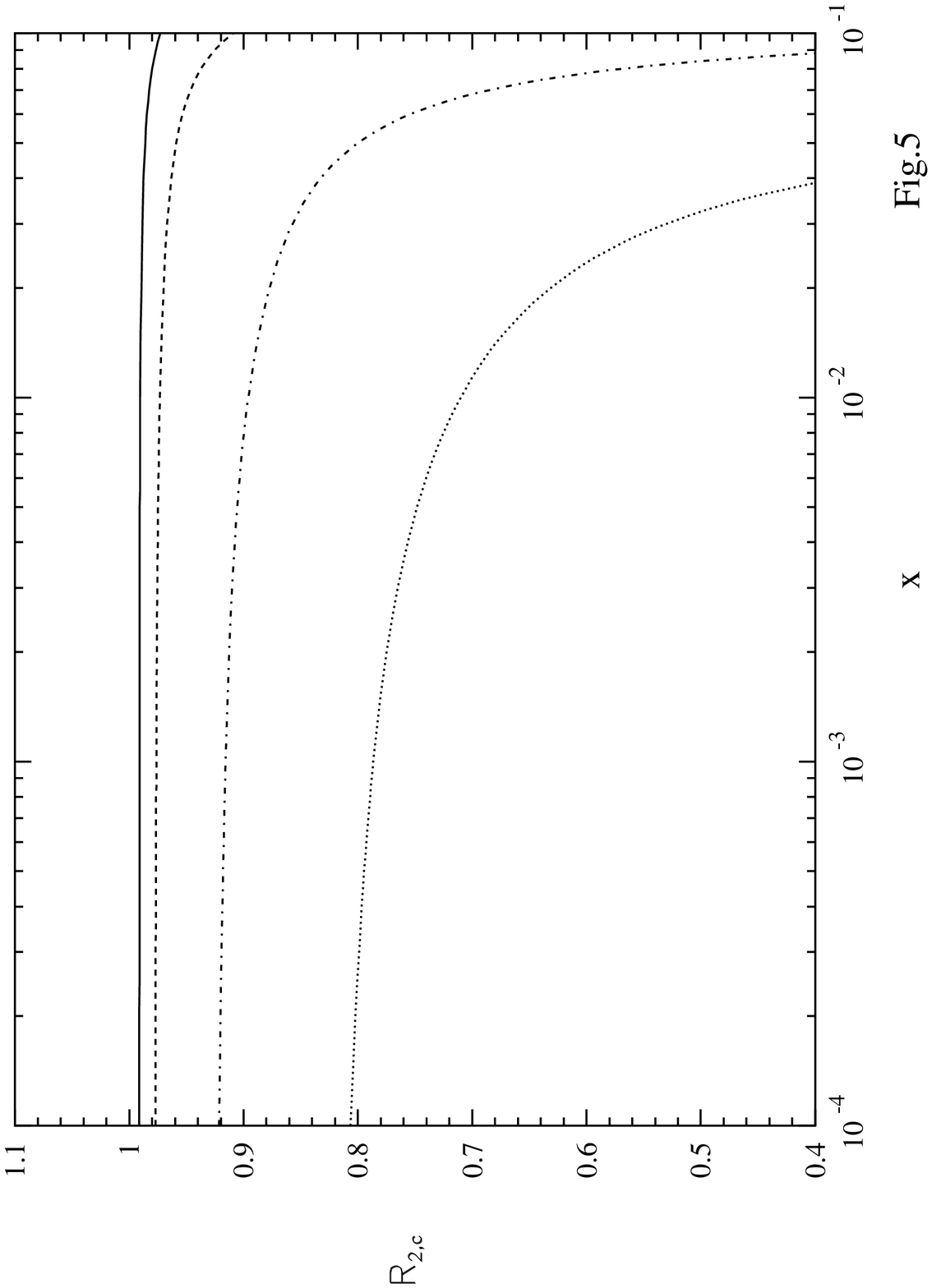}}
\end{picture}
\end{center}
\end{figure}
\begin{figure}[p]
\begin{center}
\begin{picture}(15,20)
\put(-2,0){\includegraphics{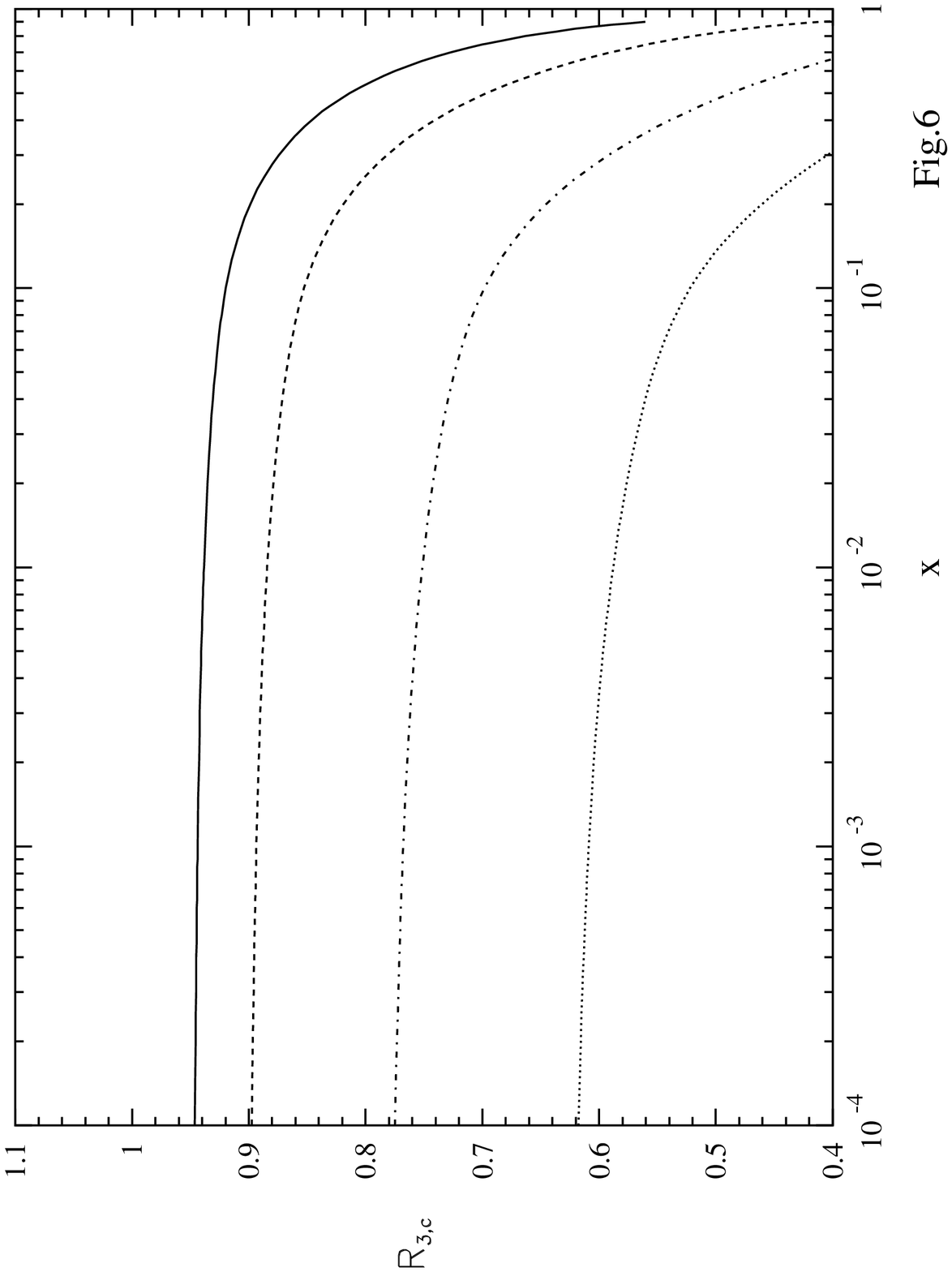}}
\end{picture}
\end{center}
\end{figure}
\begin{figure}[p]
\begin{center}
\begin{picture}(15,20)
\put(-2,0){\includegraphics{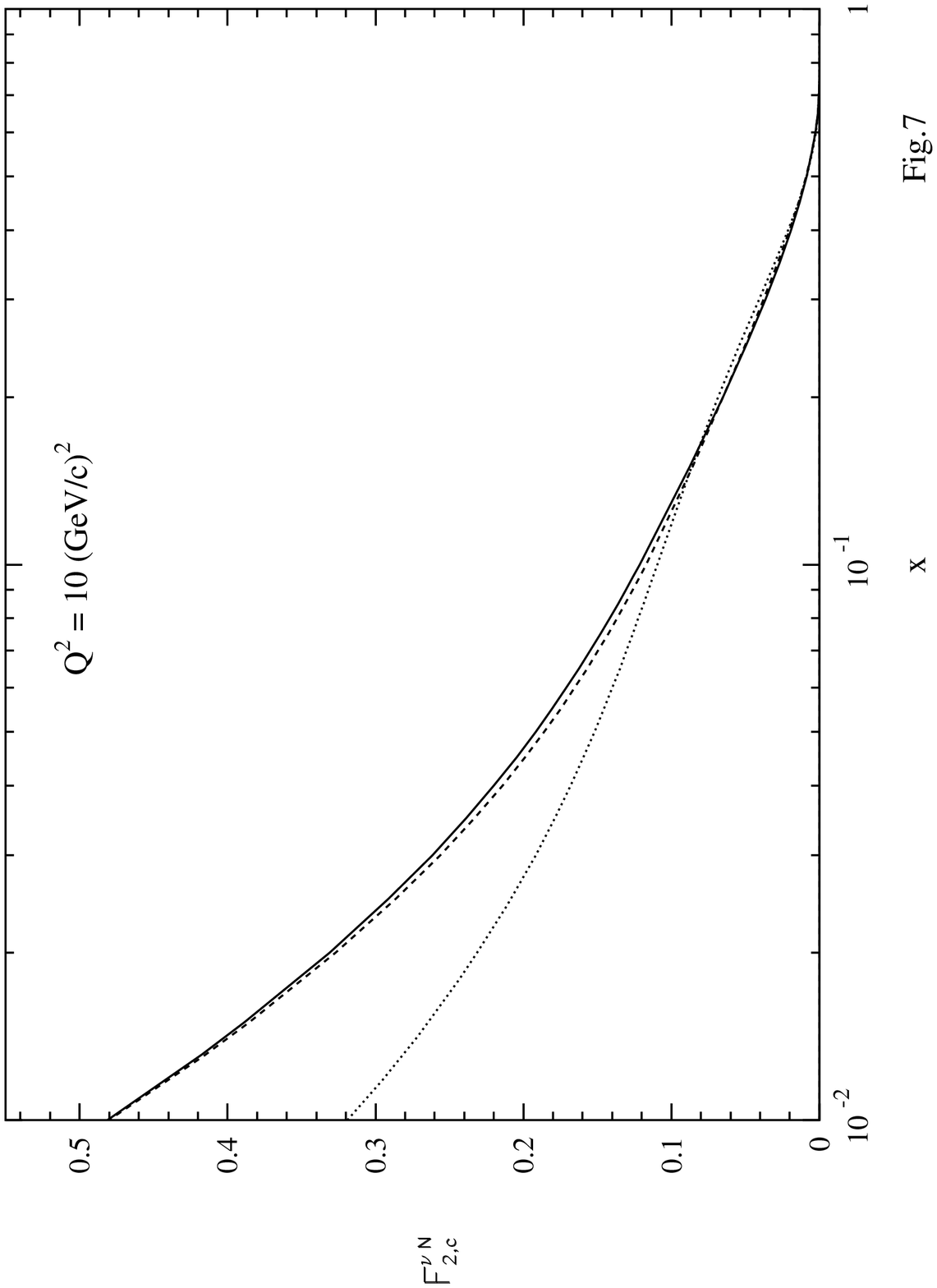}}
\end{picture}
\end{center}
\end{figure}
\begin{figure}[p]
\begin{center}
\begin{picture}(15,20)
\put(-2,0){\includegraphics{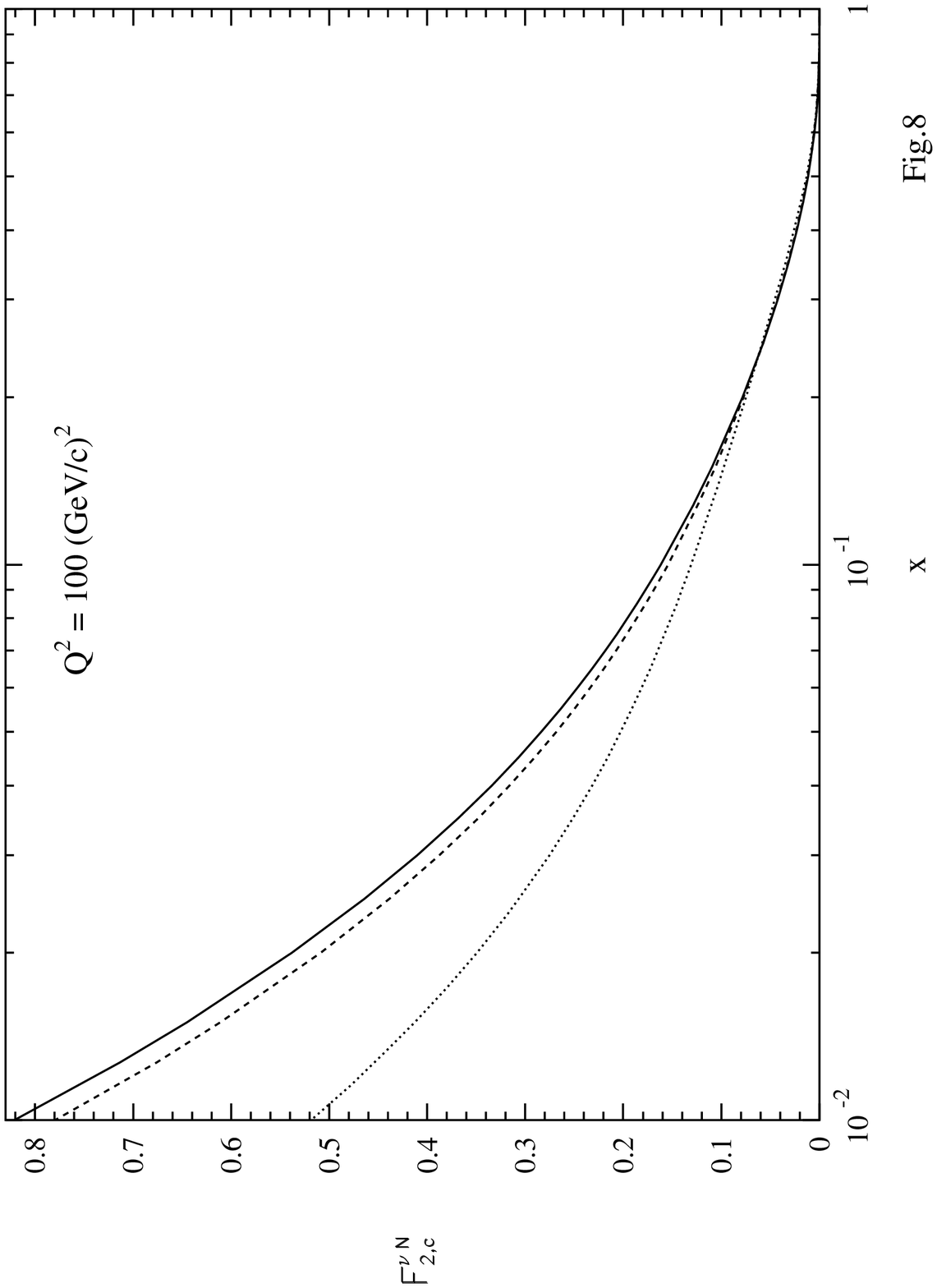}}
\end{picture}
\end{center}
\end{figure}
\begin{figure}[p]
\begin{center}
\begin{picture}(15,20)
\put(-2,0){\includegraphics{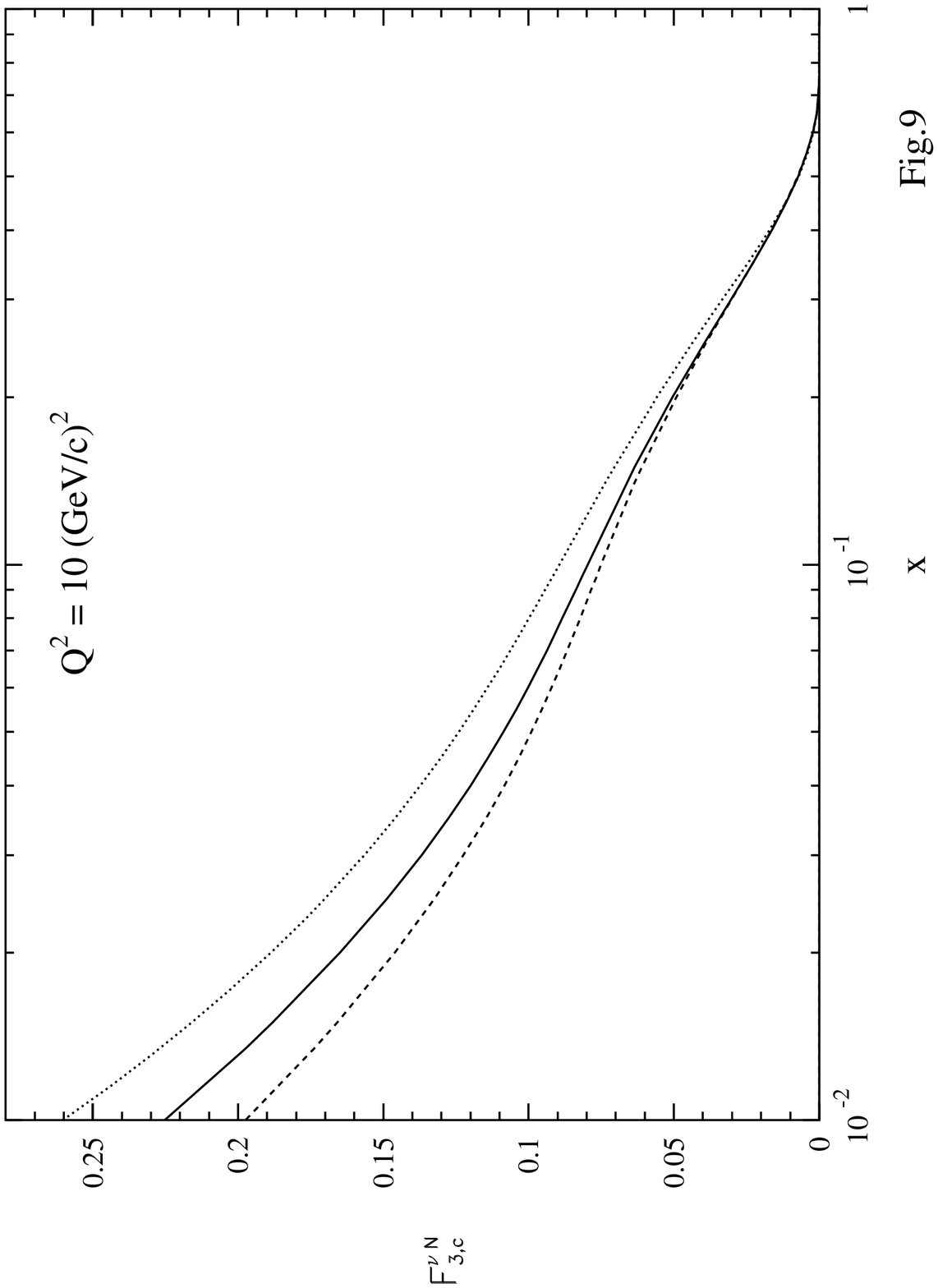}}
\end{picture}
\end{center}
\end{figure}
\begin{figure}[p]
\begin{center}
\begin{picture}(15,20)
\put(-2,0){\includegraphics{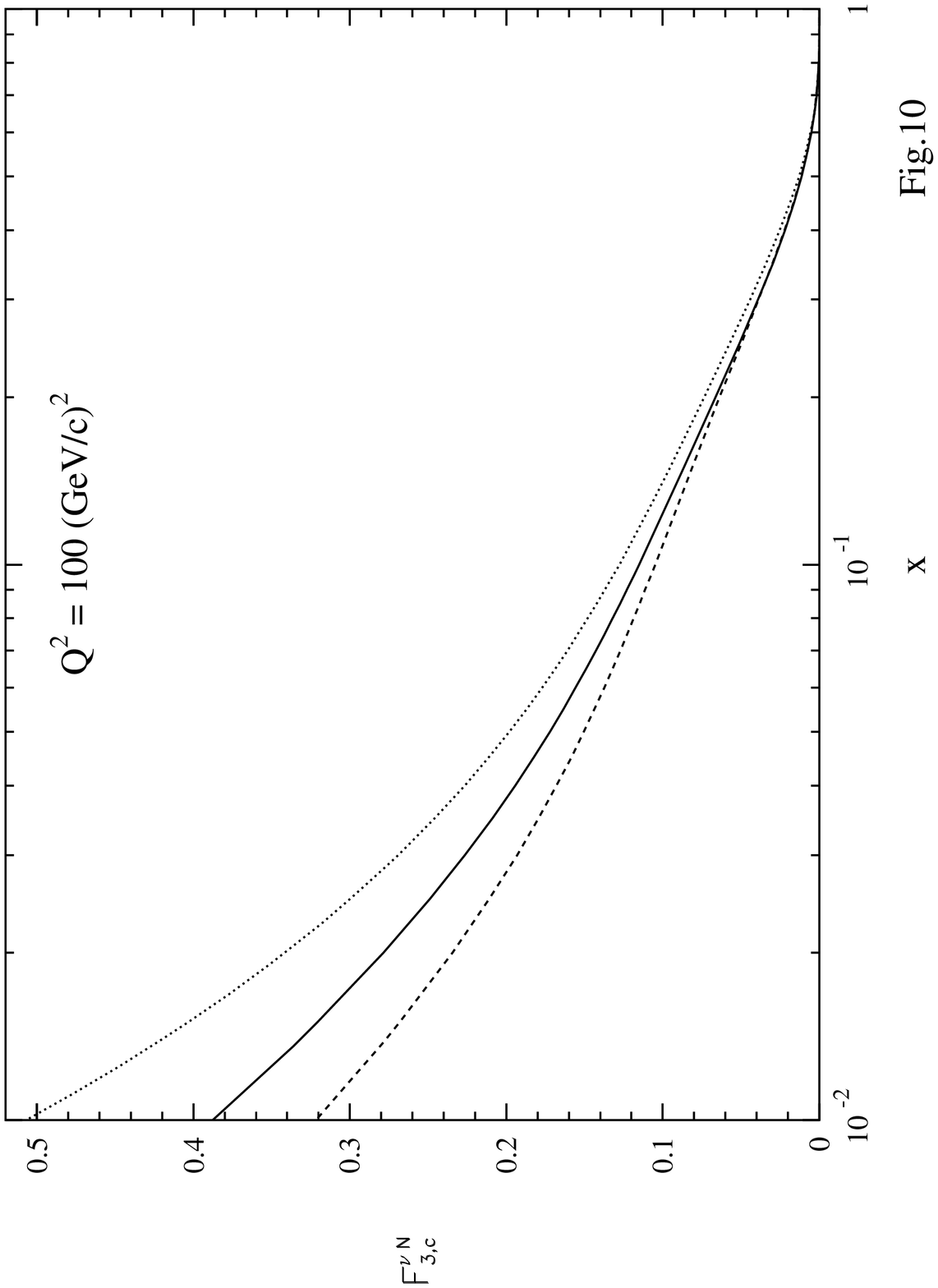}}
\end{picture}
\end{center}
\end{figure}
\begin{figure}[p]
\begin{center}
\begin{picture}(15,20)
\put(-2,0){\includegraphics{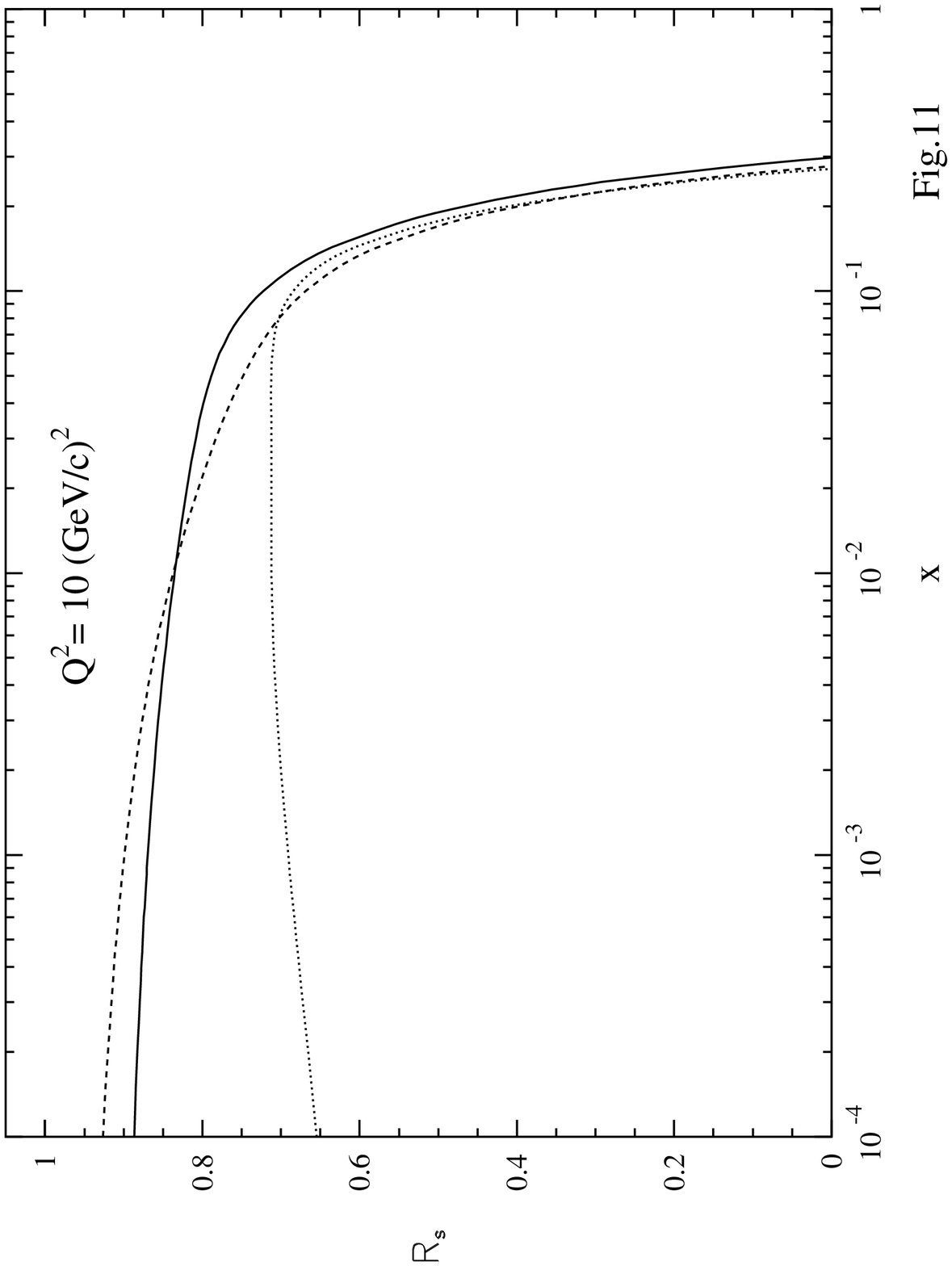}}
\end{picture}
\end{center}
\end{figure}
\begin{figure}[p]
\begin{center}
\begin{picture}(15,20)
\put(-2,0){\includegraphics{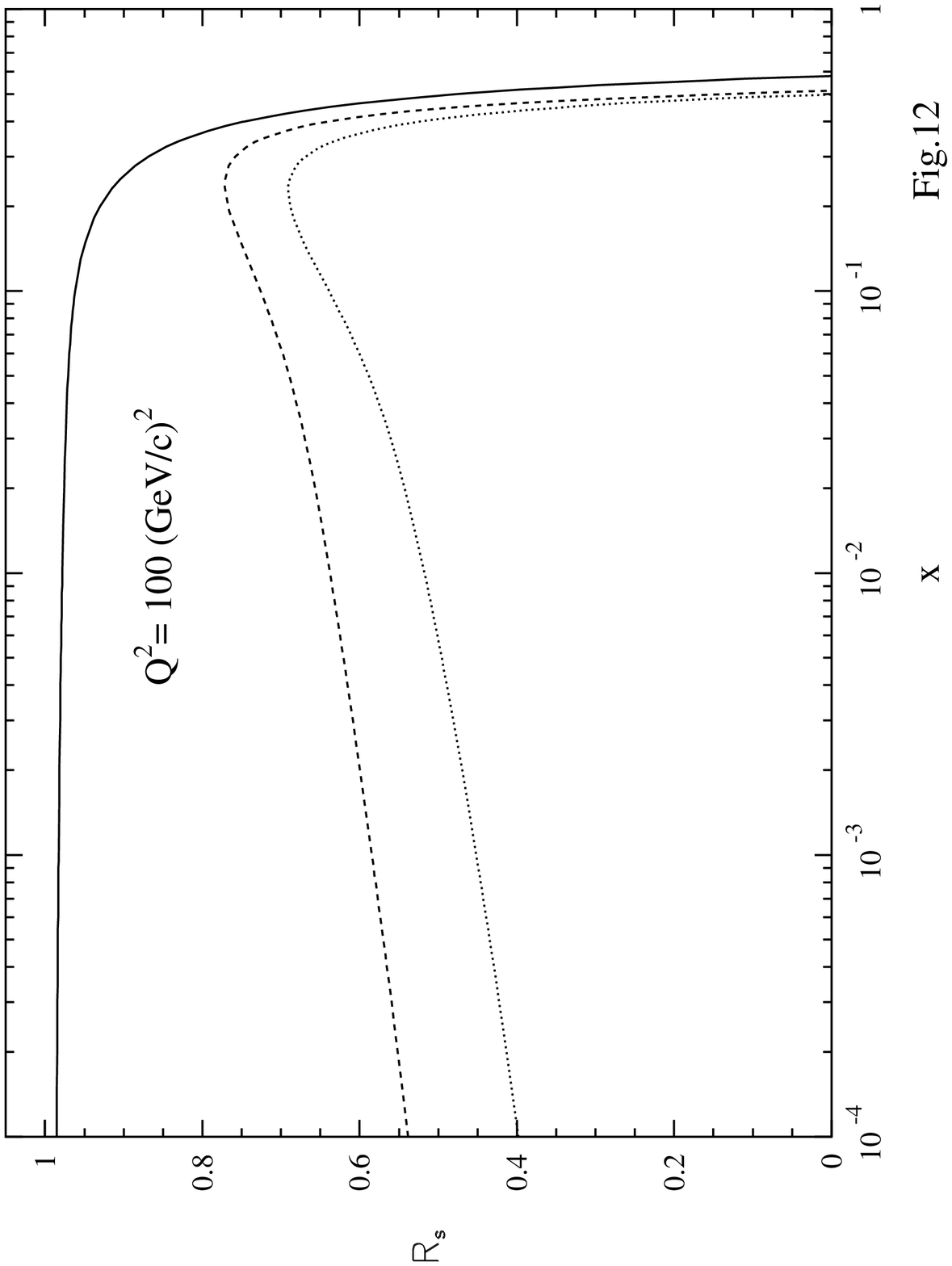}}
\end{picture}
\end{center}
\end{figure}
\begin{figure}[p]
\begin{center}
\begin{picture}(15,20)
\put(-2,0){\includegraphics{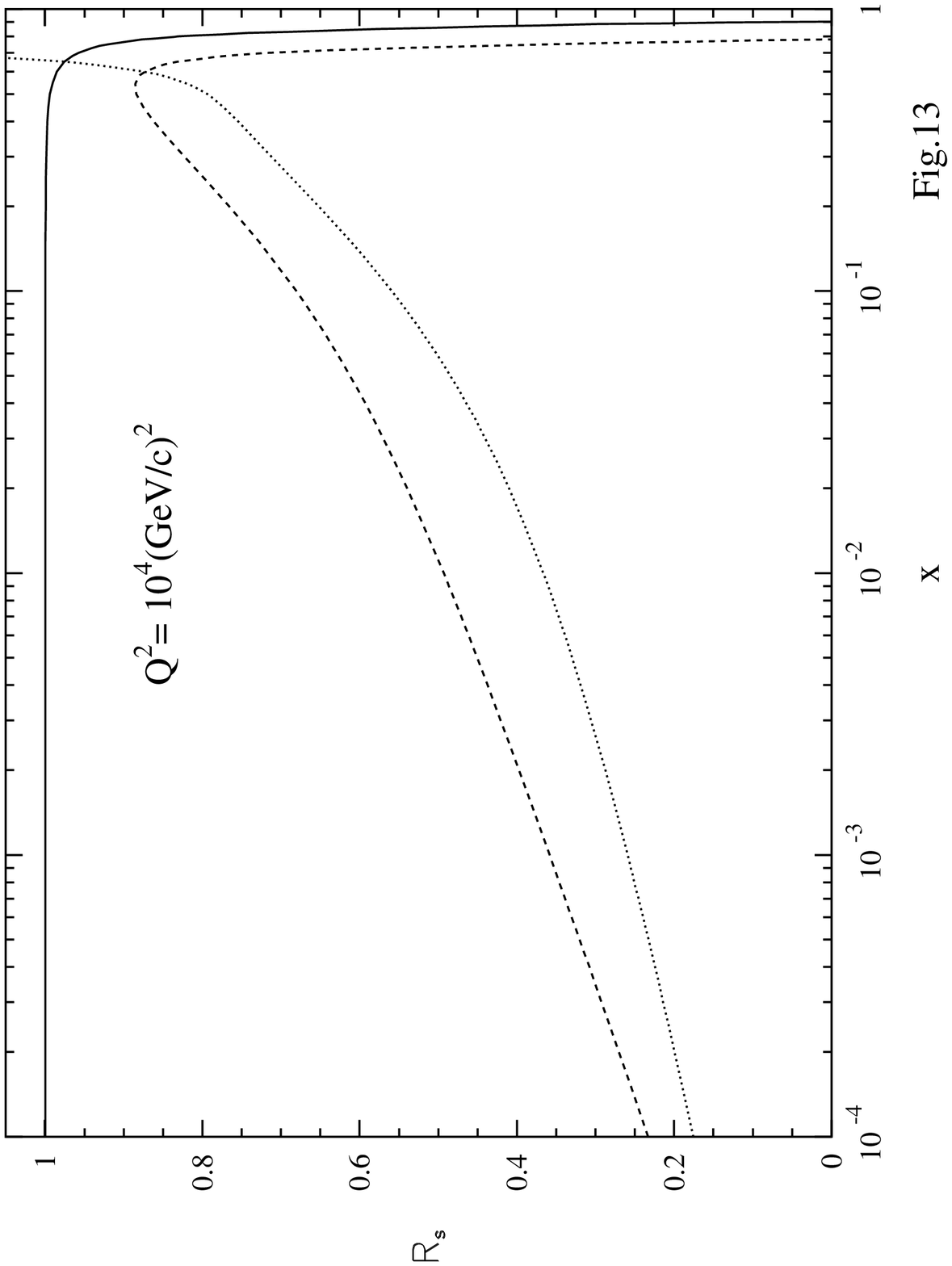}}
\end{picture}
\end{center}
\end{figure}
\begin{figure}[p]
\begin{center}
\begin{picture}(15,20)
\put(-2,0){\includegraphics{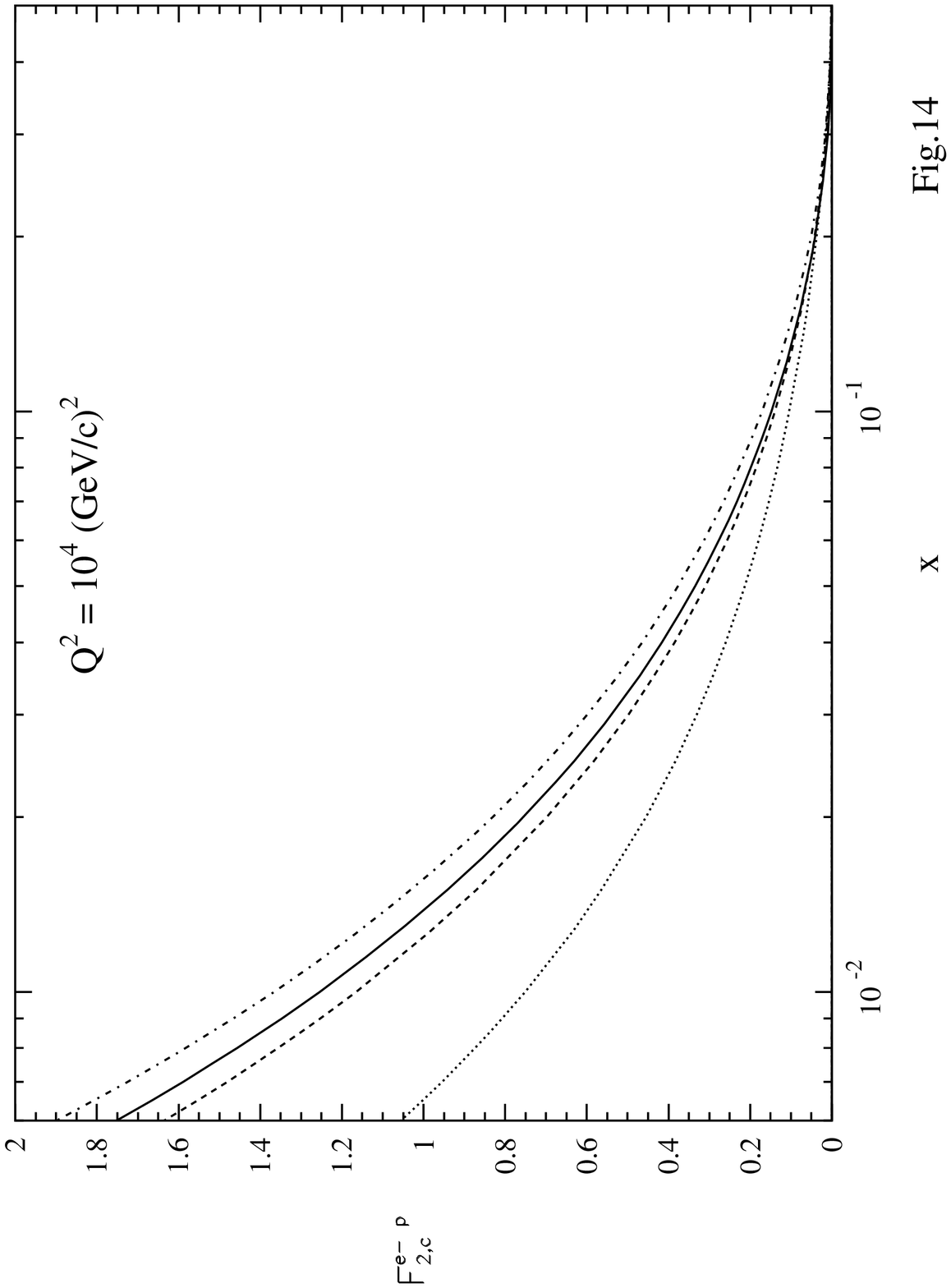}}
\end{picture}
\end{center}
\end{figure}
\begin{figure}[p]
\begin{center}
\begin{picture}(15,20)
\put(-2,0){\includegraphics{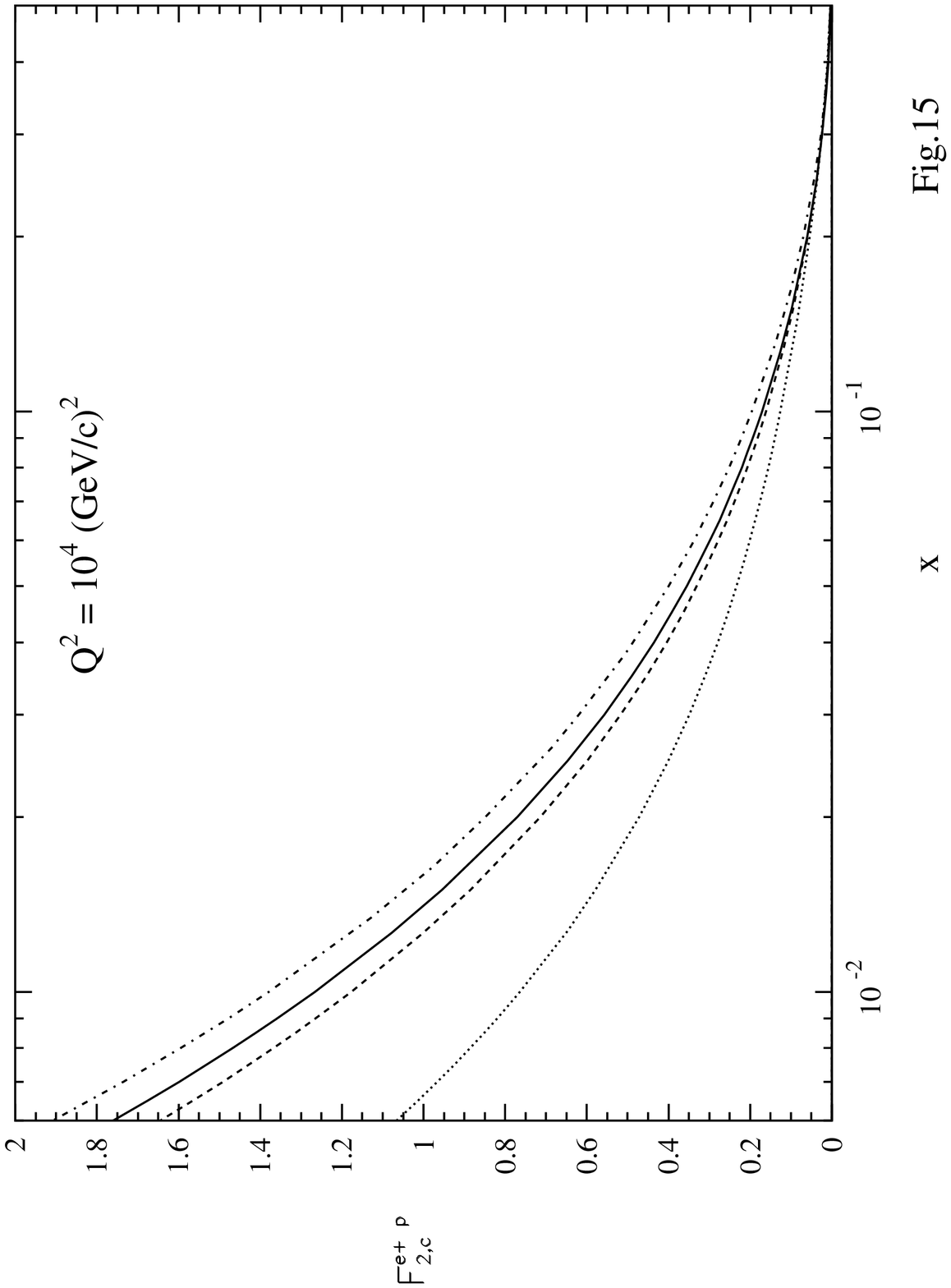}}
\end{picture}
\end{center}
\end{figure}
\begin{figure}[p]
\begin{center}
\begin{picture}(15,20)
\put(-2,0){\includegraphics{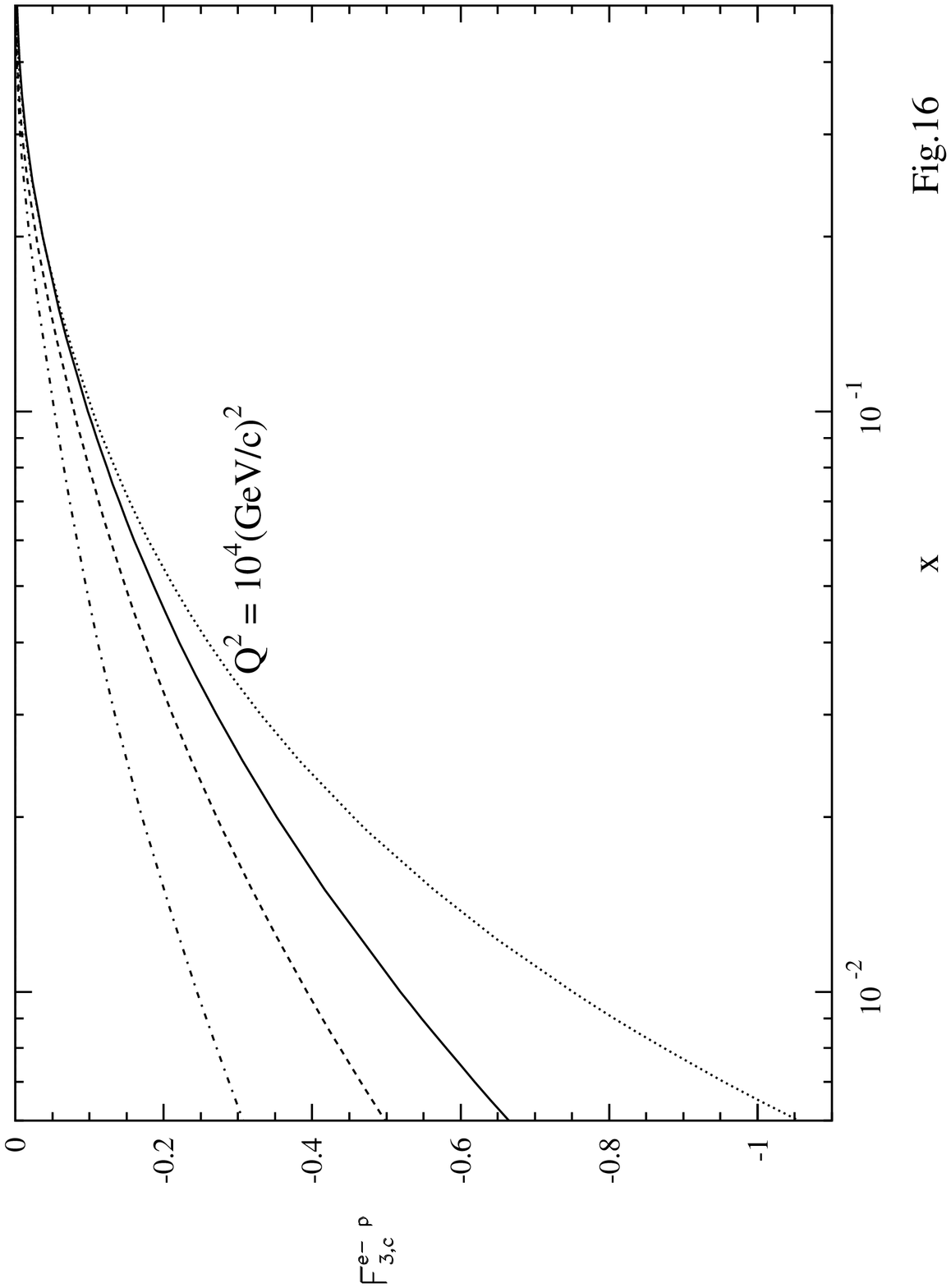}}
\end{picture}
\end{center}
\end{figure}
\begin{figure}[p]
\begin{center}
\begin{picture}(15,20)
\put(-2,0){\includegraphics{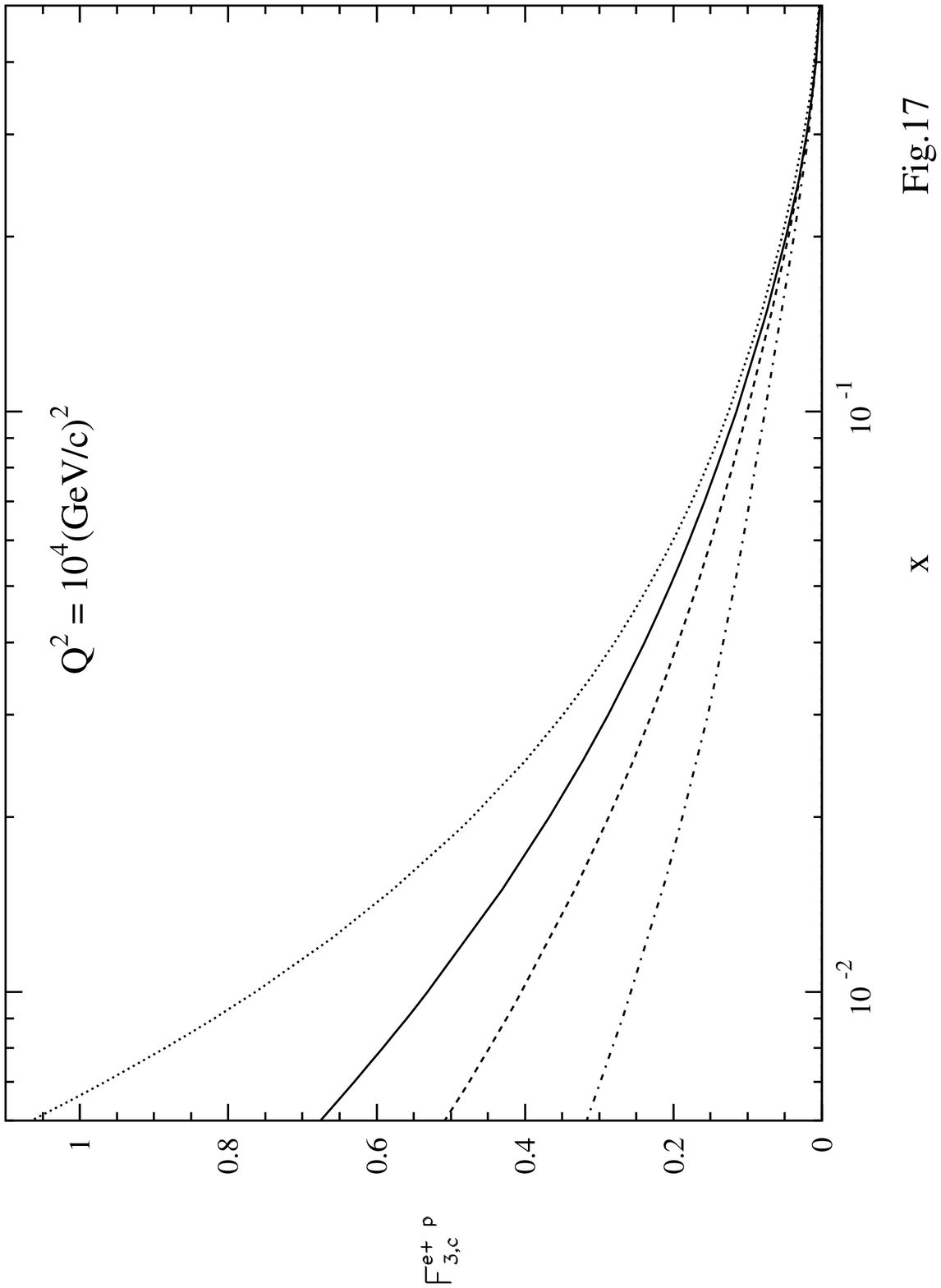}}
\end{picture}
\end{center}
\end{figure}
\end{document}

%% file: ch0.tex
\vskip 0.2cm
\hfill{NIKHEF/97-007}
\vskip 0.2cm
\hfill{INLO-PUB-2/97}
\vskip 0.2cm
\centerline{\large\bf {${\cal O}(\alpha_s^2)$ Contributions to charm production
in charged-current }}
\centerline{\large\bf {deep-inelastic lepton-hadron scattering}}
\vskip 0.2cm
\centerline {\sc M. Buza \footnote{supported by the Foundation
for Fundamental Research on Matter (FOM)}}
\centerline{\it NIKHEF/UVA,}
\centerline{\it POB 41882, NL-1009 DB Amsterdam,}
\centerline{\it The Netherlands.}
\vskip 0.2cm
\centerline {\sc W.L. van Neerven}
\centerline{\it Instituut-Lorentz,}
\centerline{\it University of Leiden,}
\centerline{\it PO Box 9506, 2300 RA Leiden,}
\centerline{\it The Netherlands.}
\vskip 0.2cm
\centerline{February 1997}
\vskip 0.5cm
\centerline{\bf Abstract}
\vskip 0.3cm
The most important part of the order $\alpha_s^2$ corrections to the charm
component of the charged-current structure functions $F_2(x,Q^2)$ and 
$F_3(x,Q^2)$ have been calculated. This calculation is based on the asymptotic 
form of the heavy-quark coefficient functions corresponding to the higher
order corrections to the
W-boson-gluon fusion process. These coefficient functions which are in
principle only valid for $Q^2 \gg m^2$ can be also used to estimate the
order $\alpha_s^2$ contributions at lower $Q^2$ values provided $x < 0.1$.
It turns out that the above corrections are appreciable in the large
$Q^2$-region and they explain the discrepancy found for the structure
functions between the fixed-flavour scheme (FFS) and the variable-flavour-number
scheme (VFNS). These corrections also hamper the extraction of the 
strange-quark density from the data obtained for the charged-current and the
electromagnetic-current processes.

\vfill
\newpage

%% file: ch1.tex
\mysection{Introduction}
\label{sec:ch1}
\newcommand{\be}{\begin{eqnarray}}
\newcommand{\ee}{\end{eqnarray}}


The study of charm production in deep-inelastic lepton-hadron
scatterings provides us with important information
about heavy-quark production mechanisms. The latter enables us to 
extract parton densities from the deep-inelastic data
in certain kinematical regions which are difficult to explore in 
other hard processes.
The best example is the gluon density $G(x,\mu^2)$ where $\mu$
denotes the factorization scale. It can be measured at small
$x$ in deep-inelastic electroproduction
$ e^{\pm} + N \rightarrow e^{\pm} + '\!\!X'$ via the 
photon-gluon fusion process 

\begin{eqnarray}
\label{eqn:41.1}
\gamma^{*} + g \rightarrow c + {\bar c}
\end{eqnarray}
where $'\!X'$ denotes any inclusive final state.
The above process dominates electroproduction of the 
charm quark if one assumes that the probability to find 
a charm quark inside the proton is zero so that the 
flavour-excitation process 

\begin{eqnarray}
\label{eqn:41.2}
\gamma^{*} + c \rightarrow c 
\end{eqnarray}
does not occur. The reactions (\ref{eqn:41.1}) and (\ref{eqn:41.2})
are sometimes also referred to as extrinsic- and intrinsic-charm 
production. In the subsequent part of this paper we will assume
that there is no charm quark in the initial state.
Another parton density which can be measured via charm production
is the strange-quark density $s(x,\mu^2)$.
The latter shows up in the charged-current process
$\nu (\bar \nu) + p \rightarrow l^{-} (l^{+}) + '\!\!X'$  ($l=e,\mu$) which
in lowest order is given by the flavour-excitation mechanism

\begin{eqnarray}
\label{eqn:41.3}
W^{+} + s \rightarrow c , \quad \quad W^{-} + {\bar s}\rightarrow {\bar c}
\, .
\end{eqnarray}
In next-to-leading order (NLO) we have the following parton subprocesses.
The first one is represented by the gluon bremsstrahlung process
\begin{eqnarray}
\label{eqn:41.4}
W^{+} + s \rightarrow c + g , \quad \quad W^{-} + {\bar s}\rightarrow {\bar c}
+g
\end{eqnarray}
which includes the one-loop corrections to the Born reaction (\ref{eqn:41.3}).
Furthermore we have the W-boson-gluon fusion mechanism given by
\begin{eqnarray}
\label{eqn:41.5}
W^{+} +  g \rightarrow c + {\bar s},
\quad \quad W^{-} + g \rightarrow {\bar c} + s
\, .
\end{eqnarray}
Although for $x \ge 0.1$ charm production is determined by flavour excitation 
it turns out  that the W-boson-gluon fusion process dominates the 
order $\alpha_s$ corrections for $x < 0.1$ and constitues a large
background for reaction (\ref{eqn:41.3}).
This means that in the region $x < 0.1$ an accurate determination of
$s(x,\mu^2)$ requires a good knowledge of $G(x,\mu^2)$.
Charm production in charged-current interactions was studied in the
experiments carried out by the CCFR-collaboration in \cite{rab},  
\cite{baz}.
Besides the strange-quark density one has also determined one of the
elements of the CKM (Cabibbo-Kobayashi-Moskawa) matrix, 
called $|V_{c d}|$, which is approximately equal to $|\sin \theta_C|$
where $\theta_C$ denotes the Cabbibo angle.
However there exists an alternative way to determine the strange-quark
density $s(x,\mu^2)$. Here the latter follows from the relation

\begin{eqnarray}
\label{eqn:41.6}
\Delta(x,\mu^2) = \frac{5}{6} F_2^{\nu N} (x,\mu^2) - 3 F_2^{\mu N}(x,\mu^2)\,.
\end{eqnarray}
Here $F_2^{\nu N}$ and $ F_2^{\mu N}$ are the structure functions
corresponding to charged-current and electromagnetic-current
induced processes respectively, where $N$ is represented by
an isoscalar nuclear target. Neglecting nuclear corrections
one obtains

\begin{eqnarray}
\label{eqn:41.7}
F_2^{l N}(x, \mu^2) 
= \frac{1}{2} \Biggl(F_2^{l p}(x, \mu^2) + F_2^{l n}(x,\mu^2)\Biggr)
\end{eqnarray}
with $l= e^{\pm }, \mu^{\pm} , 
\stackrel{(-)}{\nu_e},
\stackrel{(-)}{ \nu_{\mu}}$ and $p$ and $n$ stand
for the proton and the neutron respectively.
Further $F_2^{l N}$ denotes the full structure function where besides
charm-quark production also the other light-parton (u,d,s,g)
subprocesses are involved. If the structure functions appearing on the
right-hand side of (\ref{eqn:41.6}) are computed on the Born level
we obtain in a three-flavour scheme 

\begin{eqnarray}
\label{eqn:41.8}
\Delta(x,\mu^2) = x~s(x,\mu^2)\,.
\end{eqnarray}
In the case of a four-flavour scheme where now also the
charm quark is represented by a parton density Eq.~(\ref{eqn:41.6}) 
becomes 

\begin{eqnarray}
\label{eqn:41.9}
\Delta(x,\mu^2) = x~s(x,\mu^2) - x~c(x,\mu^2)\,.
\end{eqnarray}
The three- and four-flavour scheme are in the literature
very often referred to as fixed-flavour scheme (FFS) and variable-flavour-number
scheme (VFNS) respectively.
The structure functions $F_2^{\nu N}$ and $F_2^{\mu N}$ have been
measured by the CCFR collaboration \cite{olt}, \cite{ccfr} and
the NMC-collaboration \cite{arn}.
If one takes the data from \cite{olt}, \cite{arn} then the 
result obtained for $s(x,\mu^2)$ is quite close to the one measured in
charm production \cite{rab}, \cite{baz} (see also the discussion in 
\cite{gkr}). However there exists a discrepancy between the data of 
\cite{olt} and \cite{ccfr} which is not clarified yet. If we adopt the
three-flavour scheme for the subsequent part of our paper the
identity in (\ref{eqn:41.8}) only holds if the charm-quark mass is zero
and the QCD corrections are neglected. However as we will see later on
these corrections will considerably alter Eq.~(\ref{eqn:41.8}) in particular
in the small $x$-region. 
This is mainly due to the  W-boson-gluon fusion process in (\ref{eqn:41.5}) and
its higher order QCD corrections. The corrections due to the mass of the
charm quark become smaller when the virtuality of the W-boson
denoted by $Q^2$ will become very large.
The order $\alpha_s$ corrections given by the reactions
(\ref{eqn:41.4}), (\ref{eqn:41.5}) were calculated in \cite{got} (see also
\cite{gkr}, \cite{bo}). The dependence on the mass of the charm quark was
studied in \cite{kl} and a comparison between the FFS \cite{ghr} and VFNS
\cite{acot}  was made
in \cite{kls}, \cite{bag}. In the latter references 
the dependence of the QCD corrected 
structure functions on the factorization scale was studied too.
One of the main conclusions of these investigations is that 
the QCD corrections for $x \ge 0.1$ are small whereas for $x <0.1$ the bulk of 
the corrections is constituted by the W-boson-gluon fusion process
(\ref{eqn:41.5}). These corrections become even larger when 
$Q^2$ increases. In particular at HERA \cite{ahm}, \cite{aid}, \cite{der} where 
$200 < Q^2 < 10^4~({\rm GeV}/c)^2 $, the 
size of the contribution of the latter process 
is responsible for the discrepancy between the FFS and VFNS 
in describing the charm component
of the structure functions denoted by $F_{k,c}\,\, (k= 2,3)$ 
 as discussed in \cite{bag}.
Therefore we will concentrate ourselves in this paper on the calculation of
the order $\alpha_s^2$ contributions to $F_{k,c}$ which
are due to the order $\alpha_s$ corrections to the W-boson-gluon 
fusion process in (\ref{eqn:41.5}).This also includes the 
contributions from other heavy-quark subprocesses initiated by
the light quarks u,d and s. Unfortunately for some of them
we only have the asymptotic expressions of the heavy-quark
coefficient functions which are valid for $Q^2 \gg m^2$ where $m$ stands
for the heavy-quark mass. This implies that strictly speaking they 
are only applicable to the HERA experiments where the characteristic values for
$Q^2$ are very large. However as is shown for the electromagnetic-current
process \cite{bmsmn}-\cite{bmsn2}
, where the exact order $\alpha_s$ corrections to 
the photon-gluon fusion process are known \cite{lrsn}, the asymptotic
heavy-quark coefficient functions can be used for much lower
$Q^2$ values measured at fixed-target experiments. 
If $Q^2$ is an order of magnitude larger than $m^2$ one gets a reasonable
agreement within $10\%$ between the predictions obtained from the
exact  and asymptotic heavy-quark coefficient functions
(see \cite{bmsn1}-\cite{bmsn2}).

Our paper will be organized as follows. In Section~\ref{sec:ch2}
we present the formulae for the charged-current deep-inelastic 
structure functions expressed in convolutions of the
parton densities and the coefficient functions. 
Further we construct the asymptotic form of the order $\alpha_s^2$
contributions to the heavy-quark coefficient functions.
In Section~\ref{sec:ch3} we discuss our results and show the effect of the
latter on the charm component of the
charged-current structure functions. We also show how the
extraction of the strange density from relation (\ref{eqn:41.6}) is
influenced by the above corrections. The derivation of the asymptotic
heavy-quark coefficient functions will be presented in 
Appendix A.


%% file: ch2.tex
\mysection{${\cal O}(\alpha_s^2)$ contributions to heavy-flavour
 coefficient functions in charged-current interactions}
\label{sec:ch2}

In this section we will present the asymptotic forms of the 
heavy-quark coefficient functions up tp order $\alpha_s^2$ which
contribute to the structure functions $F_2(x,Q^2)$ and $F_3(x,Q^2)$
measured in charged-current deep-inelastic lepton-hadron scattering.
Although these coefficient functions are only valid
at $Q^2 \gg m^2$, characteristic of the values of $Q^2$ observed at HERA,
they can be also used to make a reasonable estimate of the 
charm component of the structure functions measured in fixed-target 
experiments where $Q^2$ is much smaller.
Here and in the next section we will compute all structure functions
in the so called fixed-flavour scheme (FFS).
In the case of the charm production this means that the 
constituents of the hadron are only given by the gluon and the 
three light flavours u,d,s which are described by parton densities.
The charm quark only appears in the final state of the parton subprocesses 
and its contribution to the structure functions is described
by the heavy-quark coefficient functions.
As is shown in \cite{bmsn2} the FFS is suitable to describe
the charm component to the structure functions at 
small $Q^2$-values. However at large $Q^2$ where the large
logarithms of the type $\ln^i(Q^2/m^2)\ln^j(\mu^2/m^2)$ 
dominate the heavy-quark coefficient functions this description
is not adequate anymore. In this regime it is much more useful to
adopt the  variable-flavour-number scheme (VFNS)
where now also the charm-quark contribution is described by
a parton density. This parton density represents the resummation of the
large logarithms above in all orders of perturbation theory.
We will comment on the VFNS approach in the case of charm 
production at HERA collider in the next section.
In the FFS (here three-flavour scheme) the light-parton contribution
to the structure functions measured in the process
$\nu_l + p \rightarrow l + '\!\!X'$ ($l=e,\mu$) is given by
\begin{eqnarray}
\label{eqn:42.1}
&&  F_k^{W^+ p} (x,Q^2) \equiv F_k^{\nu p} (x,Q^2)  =
\nonumber\\ && 
a_k(x) \int_x^1 \, \frac{d\, z}{z} \Biggl[ \Sigma_3\Biggl( \frac{x}{z},\mu^2
\Biggr) \tilde{\cal C}_{k,q}^{\rm PS}\Biggl(z, \frac{Q^2}{\mu^2} \Biggr)
+ G\Biggl( \frac{x}{z},\mu^2\Biggr) \tilde{\cal C}_{k,g}^{\rm S}\Biggl(z, \frac{
Q^2}{\mu^2} \Biggr)
\nonumber\\ && 
+ \Biggl\{ b_{k,\bar u} {\bar u} \Biggl( \frac{x}{z},\mu^2 \Biggr)
+ b_{k,d} \cos^2(\theta_C) d \Biggl( \frac{x}{z},\mu^2 \Biggr)
\nonumber\\ &&
+
 b_{k,s} \sin^2(\theta_C) s \Biggl( \frac{x}{z},\mu^2 \Biggr) \Biggr\}
{\cal C}_{k,q}^{\rm NS}\Biggl(z, \frac{Q^2}{\mu^2} \Biggr)
\Biggr]
\,
\end{eqnarray}
with
\begin{equation}
\label{eqn:42.2}
a_2(x) = 2 x , \quad \quad         a_3(x) = 2
\end{equation}
and for $i = u,d,s$; $\bar i = \bar u, \bar d, \bar s$ 
\begin{equation}
\label{eqn:42.3}
b_{2,i} = b_{3,i} = 1, \quad \quad  b_{2,\bar i} = - b_{3,\bar i} = 1 \, .
\end{equation}
Here $G(z,\mu^2)$ denotes the gluon density and the singlet combination
of parton densities in the three-flavour scheme is given by

\begin{eqnarray}
\label{eqn:42.4}
\Sigma_3(z,\mu^2) &=& 
u(z,\mu^2) + {\bar u} (z,\mu^2) + d (z,\mu^2) + {\bar d} (z,\mu^2)
\nonumber  \\
&& + s (z,\mu^2) + {\bar s} (z,\mu^2) \, .
\end{eqnarray}
Like the parton densities the coefficient functions 
${\cal C}_{k,i}$ ($k = 2,3; i = q,g$) can be distinguished into 
a singlet and non-singlet part indicated by the superscripts 
${\rm S}$ and ${\rm NS}$ respectively.
In (\ref{eqn:42.1}) we have split the singlet coefficient function
${\cal C}_{k,q}^{\rm S}$ in the following way
 
\begin{equation}
\label{eqn:42.5}
{\cal C}_{k,q}^{\rm S} = {\cal C}_{k,q}^{\rm NS} + {\cal C}_{k,q}^{\rm PS}
\, .
\end{equation}
The purely-singlet part of the coefficient function indicated by ${\rm PS}$
originates from quark subprocesses where the projection on the 
non-singlet channel yields zero, so that only singlet contributions remain.
They are characterized by those Feynman graphs in which only gluons are 
exchanged in the t-channel. Such graphs show up for the first time in 
order $ \alpha_s^2$.
Furthermore because of charge conjugation invariance of the 
strong interactions one has the property 

\begin{equation}
\label{eqn:42.6}
\tilde {\cal C}_{3,q}^{\rm PS} = 0, \quad \quad
\tilde {\cal C}_{3,g}^{\rm S} = 0
\, ,
\end{equation}
so that $F_3$ is determined by the non-singlet combination
of parton densities and ${\cal C}_{3,q}^{\rm NS}$ only.
In the case of $F_2$ we have extracted the overall dependence of the 
coefficient functions ${\cal C}_{2,q}^{\rm PS}$ and
${\cal C}_{2,g}^{\rm S}$ on the number of light flavours $n_f$ so that
we will denote them by $\tilde {\cal C}_{2,q}^{\rm PS}$ and 
$\tilde {\cal C}_{2,g}^{\rm S}$ respectively.
Finally, in the three-flavour scheme, the structure functions depend
on the Cabbibo angle $\theta_C$ only.

The charm contribution to the structure functions $F_k^{\nu p} \,\,(k = 2,3)$
can be written as

\begin{eqnarray}
\label{eqn:42.7}
&& \hskip -0.5cm F_{k,c}^{W^+ p} (x,Q^2,m^2)\equiv F_{k,c}^{\nu p} (x,Q^2,m^2)  =
\nonumber\\ && \hskip -0.5cm
a_k(x) \int_x^{z_{th}} \, \frac{d\, z}{z}
\Biggl[ \Sigma_3 \Biggl( \frac{x}{z},\mu^2 \Biggr)
L_{k,q}^{W,\rm PS}\Biggl(z, \frac{Q^2}{m^2}, \frac{m^2}{\mu^2} \Biggr)
+ G\Biggl( \frac{x}{z},\mu^2\Biggr)
L_{k,g}^{W,\rm S}\Biggl(z, \frac{Q^2}{m^2},\frac{m^2}{\mu^2} \Biggr)
\nonumber\\ && \hskip -0.5cm
+ \Biggl\{ b_{k,\bar u} {\bar u} \Biggl( \frac{x}{z},\mu^2 \Biggr)
+ b_{k,d} \cos^2(\theta_C) d \Biggl( \frac{x}{z},\mu^2 \Biggr)
+ b_{k,s} \sin^2(\theta_C) s \Biggl( \frac{x}{z},\mu^2 \Biggr) \Biggr\}
\nonumber\\ && \hskip -0.5cm
\times
L_{k,q}^{W,\rm NS}\Biggl(z, \frac{Q^2}{m^2}, \frac{m^2}{\mu^2} \Biggr)
\Biggr]
+ a_k(x) \int_x^{z_{th}'} \, \frac{d\, z}{z}
 \Biggl[ \Sigma_3 \Biggl( \frac{x}{z},\mu^2 \Biggr)
H_{k,q}^{W,\rm PS}\Biggl(z, \frac{Q^2}{m^2}, \frac{m^2}{\mu^2} \Biggr)
\nonumber\\ && \hskip -0.5cm
+G\Biggl( \frac{x}{z},\mu^2\Biggr)
H_{k,g}^{W,\rm S}\Biggl(z, \frac{Q^2}{m^2},\frac{m^2}{\mu^2} \Biggr)
+ \Biggl\{ b_{k,d} \sin^2(\theta_C) d \Biggl( \frac{x}{z},\mu^2 \Biggr)
\nonumber\\ && \hskip -0.5cm
+ b_{k,s}\cos^2(\theta_C) s \Biggl( \frac{x}{z},\mu^2 \Biggr) \Biggr\}
H_{k,q}^{W,\rm NS}\Biggl(z, \frac{Q^2}{m^2}, \frac{m^2}{\mu^2} \Biggr)
\Biggr]
\,
\end{eqnarray}
with
\begin{eqnarray}
\label{eqn:42.8}
z_{th} = \frac{Q^2}{Q^2 + 4 m^2},\quad \quad z_{th}' = \frac{Q^2}{Q^2 + m^2}
\, .
\end{eqnarray}
The above upper boundaries of the integrals in (\ref{eqn:42.7})
follow from the threshold conditions. In the first part
of Eq.~(\ref{eqn:42.7}) the heavy-quark coefficient functions
$L_{k,i}^{W}$ originate from parton subprocesses with a heavy quark
and a heavy anti-quark in the final state. Here the W-boson is attached to the
light quarks u,d,s only. Hence the partonic centre of mass energy squared 
$\hat{s}$ has to satisfy the condition $\hat{s} \geq 4 m^2$ with
$\hat{s}=Q^2 (1 - z)/z$. In the second part of (\ref{eqn:42.7})
the heavy-quark coefficient functions $H_{k,i}^{W}$
originate from  the parton subprocesses where either the heavy quark
or the heavy anti-quark appears in the final state which is 
accompanied by a light (anti-) quark. 
Furthermore the W-boson also couples to the heavy (anti-) quark.
Therefore $\hat{s} \geq  m^2$.
Notice that in (\ref{eqn:42.7}) we have limited ourselves to those
parton subprocesses which have no more than two heavy (anti-) quarks
in the final state. 
The notation in (\ref{eqn:42.7}) is the same as given for the light-parton
contributions to the structure functions listed below (\ref{eqn:42.1}).
Because of charge conjugation invariance of the strong interactions the
$L_{k,i}^{W}$ satisfy similar relations as quoted in (\ref{eqn:42.6})
i.e. 

\begin{eqnarray}
\label{eqn:42.9}
L_{3,q}^{W,\rm PS} = 0, \quad \quad L_{3,g}^{W,\rm S} = 0
\, .
\end{eqnarray}

The structure functions $F_k^{\bar{\nu} p}$ and $F_{k,c}^{\bar{\nu} p}$
which show up in the anti-neutrino reaction 
$\bar{\nu}_l + p \rightarrow \bar{l} + '\!\!X'$  
can be derived from Eqs.~(\ref{eqn:42.1}) and (\ref{eqn:42.7}) by
the replacements $u  \leftrightarrow \bar{u}$, 
$d  \leftrightarrow \bar{d}$, $s \leftrightarrow \bar{s}$.
Moreover in the case of $F_{3,c}^{\bar{\nu} p}$
the heavy-quark coefficient functions $H_{3,q}^{W,\rm PS}$
and  $H_{3,g}^{W,\rm S}$ get  a relative minus sign with respect to 
the ones showing up in $F_{3,c}^{\nu p}$ (\ref{eqn:42.7}).

In the case of an isoscalar target one also has to compute the
structure functions for the neutron given by $F_k^{\nu n}$,
$F_k^{\bar{\nu} n}$, $F_{k,c}^{\nu n}$, $F_{k,c}^{\bar{\nu} n}$.
The latter can be derived from the proton structure functions
by the replacements $u \leftrightarrow d$ and 
$\bar{u} \leftrightarrow \bar{d}$.

In the next section we want to compute the structure
functions in (\ref{eqn:42.1}) and (\ref{eqn:42.7}) up
to next-to-leading order (NLO).
Moreover we want to include the order $\alpha_s^2$ contributions
to the heavy-quark coefficient functions 
$L_{k,q}^{W,\rm NS}$, $H_{k,q}^{W,\rm PS}$ and $H_{k,g}^{W,\rm S}$
appearing in $F_{k,c}^{\nu p}$ (\ref{eqn:42.7}).
Notice that up to order $\alpha_s^2$ the coefficient functions
$L_{2,q}^{W,\rm PS}$ and  $L_{2,g}^{W,\rm S}$ do not contribute
because the corresponding parton subprocesses are of the order
$\alpha_s^4$ and $\alpha_s^3$ respectively. For $L_{3,q}^{W,\rm PS}$ and  
$L_{3,g}^{W,\rm S}$ see (\ref{eqn:42.9}).
Furthermore we do not include the order $\alpha_s^2$ contributions to
$H_{k,q}^{W,\rm NS}$. The latter coefficient function is determined by
the higher order QCD corrections to the flavour-excitation process
(\ref{eqn:41.3}). In order $\alpha_s$ it 
turns out that these corrections are much smaller than those 
originating from the W-boson-gluon fusion process
(\ref{eqn:41.5}) leading to the coefficient function
$H_{k,g}^{W,\rm S}$.
Hence we expect that beyond order $\alpha_s$ the radiative corrections are
dominated by the W-boson-gluon fusion mechanism which is also
indirectly present in the parton subprocess contributing to
$H_{k,q}^{W,\rm PS}$.

For the computation of $F_{k}^{\nu p}$, $F_{k}^{\bar{\nu} p}$ (\ref{eqn:42.1})
one needs the next-to-leading-log (NLL)  parton densities and 
the light-parton coefficient functions ${\cal C}_{k,i}$ $(k=2,3; i= q,g)$
corrected up to order $\alpha_s$. Notice that 
$\tilde {\cal C}_{2,q}^{\rm PS}$ vanishes up to this order.
The representation in the $\overline{\rm MS}$-scheme can be found in Appendix
 I of \cite{fp}. For the calculation of 
$F_{k,c}^{\nu p}$, $F_{k,c}^{\bar{\nu} p}$ 
the heavy-quark coefficient functions $H_{k,q}^{W,\rm NS}$
and $H_{k,g}^{W,\rm S}$ have been computed in \cite{got} up to order $\alpha_s$
(see also \cite{gkr}, \cite{bo}). The latter are represented in the
$\overline{\rm MS}$-scheme in Appendix A of \cite{gkr}.
In this paper we want to compute the order $\alpha_s^2$
contributions to $L_{k,q}^{W,\rm NS}$, $H_{k,q}^{W,\rm PS}$ and 
$H_{k,g}^{W,\rm S}$. The coefficient function $L_{k,q}^{W,\rm NS}$ is 
given by the Compton reaction

\begin{eqnarray}
\label{eqn:42.10}
W + q(\bar{q}) \rightarrow q'( \bar{q}') + c + \bar{c}
\end{eqnarray}
where the heavy-quark pair originates from gluon splitting and the gluon
is radiated off the light quarks q and q' ($q,q'=u,d,s$).
The reaction above
also includes the two-loop vertex correction
to the subprocess $W + q (\bar{q}) \rightarrow q' (\bar{q}')$
containing the charm-loop contribution to the gluon self-energy.
This we do in order to improve the large $Q^2$ behaviour of the
coefficient function. Without this vertex correction 
$L_{k,q}^{W,\rm NS} \sim \ln^3 (Q^2/m^2) $ whereas it behaves like
$\ln^2 (Q^2/m^2)$ if this correction has been included.
The coefficient function $H_{k,q}^{W,\rm PS}$ is computed from the
Bethe-Heitler process given by the reaction

\begin{eqnarray}
\label{eqn:42.11}
W + q (\bar{q}) \rightarrow q (\bar{q})+ c(\bar{c}) + \bar{q}' (q')
\end{eqnarray}
with $q= u,d,s$ and $q'=d,s$. Like 
reaction (\ref{eqn:42.10}) process (\ref{eqn:42.11})
shows up for the first time if the QCD corrections are
calculated up to order $\alpha_s^2$. 
Contrary to (\ref{eqn:42.10}), where the W-boson is only attached to the
light quarks q and q', the vector boson in reaction
(\ref{eqn:42.11}) is now coupled to the charm quark and
the light quark q'.
Finally we have the W-boson-gluon fusion mechanism
 which contributes to $H_{k,g}^{W, \rm S}$. In order 
$\alpha_s$ (lowest order) this production mechanism
is given by the process (see also (\ref{eqn:41.5}))

\begin{eqnarray}
\label{eqn:42.12}
W + g \rightarrow c (\bar{c}) + \bar{q}' (q')
\end{eqnarray}
with $q' = d,s$. In NLO we have to include all virtual
corrections to reaction (\ref{eqn:42.12}) and to add
the contributions due to the gluon bremsstrahlung process

\begin{eqnarray}
\label{eqn:42.13}
W + g \rightarrow c (\bar{c}) + \bar{q}' (q') + g \, .
\end{eqnarray}

Contrary to $L_{k,q}^{W,\rm NS}$ for which the exact order
$\alpha_s^2$ expressions exist, see \cite{bmsmn} ($k=2$) and \cite{bmsn1} 
($k=3$) we do not have
the exact form of $H_{k,q}^{W,\rm PS}$ and $H_{k,g}^{W,\rm S}$ in the
same order.
This is in contrast to the heavy-flavour electroproduction where the
deep-inelastic process only proceeds via the exchange of a virtual photon.
In this case the exact order $\alpha_s^2$ contributions to the
heavy-quark coefficients $H_{2,q}^{\gamma ,\rm PS}$ and 
$H_{2,q}^{\gamma ,\rm S}$
are known and they can be found in \cite{bmsmn}. For the charged-current 
process the calculation of these coefficient functions is still more
difficult than the ones for the electromagnetic-current reaction as
given above. This can be attributed to the mass of the light quarks q' in
(\ref{eqn:42.11})-(\ref{eqn:42.13}) which is usually put to be
equal to zero. Therefore the power of the collinear singularities,
occurring in the higher order radiative corrections to the charged-current 
reaction,
is much higher than the one appearing in the electromagnetic-current process 
where also q' stands for the charm quark. This in particular will 
complicate the calculation of the many particle phase space integrals.
The latter are even more complicated than those encountered in
parton subprocesses where all masses of the particles in the final state
are equal to zero. These type of phase space integrals have
been calculated in \cite{zn}.
However these complications can be avoided if one is only interested in
the asymptotic form of the heavy-quark coefficient functions given for
$Q^2 \gg m^2$. This form can be derived using the renormalization
group and mass-factorization techniques. 
These methods have been applied to obtain the heavy-quark coefficient
functions in the limit $Q^2 \gg m^2$ for unpolarized \cite{bmsmn}
and polarized \cite{bmsn1} electromagnetic-current deep-inelastic lepton-hadron
scattering processes. Here in the FFS the light parton contribution to the
structure function $F_{2}^{l p}(x,Q^2)$ ($l=e,\mu$) is given by
\begin{eqnarray}
\label{eqn:42.14}
&& \hskip -1.5cm  F_2^{\gamma p}(x ,Q^2) \equiv F_2^{l p}(x, Q^2)
 = x \int_x^1 \frac{d \, z}{z}
\Biggl[ \frac{2}{3} \Biggl\{
\Sigma_3 \Biggl(\frac{x}{z},\mu^2\Biggr)
\tilde {\cal C}_{2,q}^{\rm PS} (z, \mu^2)
\nonumber   \\ && \hskip -1.5cm
+ G\Biggl(\frac{x}{z},\mu^2\Biggr)
\tilde {\cal C}_{2,g}^{\rm S} (z, \mu^2)
\Biggr\}
+ \Biggl\{ \frac{4}{9} \Biggl[ u \Biggl(\frac{x}{z},\mu^2\Biggr)
+ \bar{u}\Biggl(\frac{x}{z},\mu^2\Biggr)\Biggr]
\nonumber \\ && \hskip -1.5cm
+ \frac{1}{9}  \Biggl[ d\Biggl(\frac{x}{z},\mu^2\Biggr)
+ \bar{d}\Biggl(\frac{x}{z},\mu^2\Biggr)
+ s\Biggl(\frac{x}{z},\mu^2\Biggr)
+ \bar{s}\Biggl(\frac{x}{z},\mu^2\Biggr)\Biggr]
\Biggr\}
{\cal C}_{2,q}^{\rm NS}(z, \mu^2) \Biggr]
\end{eqnarray}
and the charm component reads
\begin{eqnarray}
\label{eqn:42.15}
&&  F_{2,c}^{\gamma p}(x ,Q^2,m^2) \equiv F_{2,c}^{l p}(x, Q^2,m^2)
%
 = x \int_x^{z_{th}}
\frac{d \, z}{z}
\Biggl[ \frac{2}{3} \Biggl\{
\Sigma_3 \Biggl(\frac{x}{z},\mu^2\Biggr)
\nonumber \\ &&
\times
L_{2,q}^{\gamma, \rm PS} \Biggl(z, \frac{Q^2}{m^2}, \frac{m^2}{\mu^2}\Biggr)
+ G\Biggl(\frac{x}{z},\mu^2\Biggr)
L_{2,g}^{\gamma, \rm S} \Biggl(z, \frac{Q^2}{m^2}, \frac{m^2}{\mu^2}\Biggr)
\Biggr\}
+ \Biggl\{ \frac{4}{9}
\Biggl[ u \Biggl(\frac{x}{z},\mu^2\Biggr)
\nonumber \\ &&
+ \bar{u}\Biggl(\frac{x}{z},\mu^2\Biggr)\Biggr]
%
+ \frac{1}{9}  \Biggl[ d\Biggl(\frac{x}{z},\mu^2\Biggr)
+ \bar{d}\Biggl(\frac{x}{z},\mu^2\Biggr)
+ s\Biggl(\frac{x}{z},\mu^2\Biggr)
+ \bar{s}\Biggl(\frac{x}{z},\mu^2\Biggr)\Biggr]
\Biggr\}
\nonumber \\ &&
\times
L_{2,q}^{\gamma, \rm NS} \Biggl(z, \frac{Q^2}{m^2}, \frac{m^2}{\mu^2}\Biggr)
\Biggr]
+\frac{4}{9} x \int_x^{z_{th}}
\frac{d \, z}{z}
\Biggl[
\Sigma_3 \Biggl(\frac{x}{z},\mu^2\Biggr)
H_{2,q}^{\gamma, \rm PS} \Biggl(z, \frac{Q^2}{m^2}, \frac{m^2}{\mu^2}\Biggr)
\nonumber \\ &&
+ G\Biggl(\frac{x}{z},\mu^2\Biggr)
H_{2,g}^{\gamma, \rm S} \Biggl(z, \frac{Q^2}{m^2}, \frac{m^2}{\mu^2}\Biggr)
\Biggr]
\,.
\end{eqnarray}
The notations above are the same as those given for the charged-current
structure functions in (\ref{eqn:42.1}) and (\ref{eqn:42.7}). Notice
that the light-parton coefficient functions ${\cal C}_{k,i}$ are the same
in both reactions which means that they do not depend on the probe $W$ or
$\gamma$. This does not hold for the heavy-quark coefficient functions.
The most important result of these 
calculations is that the 
$F_{2,c}^{l p}(x,Q^2)$ does not change when the exact heavy-quark 
coefficient functions are replaced by their asymptotic expressions 
provided $Q^2 >  Q_{min}^2$. In the case of charm production
$Q_{min}^2 = 20~({\rm GeV}/c)^2$ for $x < 0.1$.
Furthermore it turns out that the value obtained for $Q_{min}^2$ 
is not altered while going  form the LO to the NLO approximation.
If we now assume that the same property also holds for the charged-current
structure functions $F_{k}^{\nu p}(x, Q^2)$, $F_{k}^{\bar{\nu} p}(x, Q^2)$
then the asymptotic forms of $H_{k,i}^{W}$ can be used to make fairly
good predictions of the order $\alpha_s^2$ contributions. These 
predictions will become very accurate for experiments carried out at HERA since
the $Q^2$-values are very large and the $x$-values are reasonable small i.e.
$200 < Q^2 < 10^4~({\rm GeV}/c)^2$ and 
$0.006 < x < 0.5$.
Even for fixed-target experiments where the characteristic 
$Q^2$-values are much smaller one can make reliable
predictions unless one enters the threshold region where $Q^2$ is too small
$(Q^2 < 20~({\rm GeV}/c)^2)$ and/or $x$ is too large $(x > 0.1)$.

The asymptotic heavy-quark coefficient functions $H_{k,i}^{W}$ for
the charged-current process are derived in Appendix A. It appears
that they can be expressed into the asymptotic heavy-quark coefficient functions
$H_{k,i}^{\gamma}$ in (\ref{eqn:42.15}) 
computed for the electromagnetic-current
process (see Appendix D of \cite{bmsmn}).
Let us first first expand these coefficient functions in a power
series like 

\begin{eqnarray}
\label{eqn:42.16}
H_{k,i}^{V} = \sum_{l=1}^{\infty} \Biggl( \frac{\alpha_s}{4 \pi} \Biggr)^l
H_{k,i}^{V,(l)}
\end{eqnarray}
with $V=\gamma,W; k = 2,3$ and $i=q,g$.
{}From Appendix A it follows that the coefficients $H_{k,i}^{V,(l)}$
up to order $\alpha_s^2$ in the limit $Q^2 \gg m^2$ are related to each
other as follows

\begin{eqnarray}
\label{eqn:42.17}
H_{k,g}^{W,\rm S,(1)}\Biggl(z, \frac{Q^2}{m^2},\frac{m^2}{\mu^2} \Biggr) = 
\frac{1}{2} 
H_{2,g}^{\gamma,\rm S,(1)}\Biggl(z, \frac{Q^2}{m^2},\frac{m^2}{\mu^2} \Biggr)
+\frac{1}{2}
\epsilon_k \tilde {\cal C}_{2,g}^{\rm S,(1)} \Biggl(z, \frac{Q^2}{\mu^2} \Biggr)
\, ,
\end{eqnarray}

\begin{eqnarray}
\label{eqn:42.18}
&& H_{k,g}^{W,\rm S,(2)}\Biggl(z, \frac{Q^2}{m^2},\frac{m^2}{\mu^2} \Biggr) =
\frac{1}{2}
H_{2,g}^{\gamma,\rm S,(2)}\Biggl(z, \frac{Q^2}{m^2},\frac{m^2}{\mu^2} \Biggr)
+\frac{1}{2}
\epsilon_k \tilde {\cal C}_{2,g}^{\rm S,(2)} \Biggl(z, \frac{Q^2}{\mu^2} \Biggr)
\nonumber \\ &&
+ \frac{1}{2}
\int_{z}^{1} \frac{d \, y}{y}
A_{cg}^{\rm S,(1)} \Biggl(\frac{z}{y},  \frac{m^2}{\mu^2} \Biggr)
\Biggl\{ {\cal C}_{k,q}^{\rm S,(1)} \Biggl(z, \frac{Q^2}{\mu^2} \Biggr)
-  {\cal C}_{2,q}^{\rm S,(1)} \Biggl(z, \frac{Q^2}{\mu^2} \Biggr)
\Biggr\}
\, ,
\end{eqnarray}

\begin{eqnarray}
\label{eqn:42.19}
&& \hskip -2.0cm
H_{k,q}^{W,\rm PS,(2)} \Biggl(z, \frac{Q^2}{m^2},\frac{m^2}{\mu^2} \Biggr) =
\frac{1}{2}
H_{2,q}^{\gamma,\rm PS,(2)} \Biggl(z, \frac{Q^2}{m^2},\frac{m^2}{\mu^2} \Biggr)
+\frac{1}{2}
\epsilon_k \tilde {\cal C}_{2,q}^{\rm PS, (2)} \Biggl(z, \frac{Q^2}{\mu^2}\Biggr
)
\,
\end{eqnarray}
with 
\begin{eqnarray}
\label{eqn:42.20}
\epsilon_2 = 1,  \quad \quad \epsilon_3 = -1
\end{eqnarray}
where $H_{2,i}^{\gamma,(l)} \,\, (i = q,g)$ can be found in Appendix D of 
\cite{bmsmn}. Further $A_{cg}^{\rm S,(1)}$ stands for the one-loop
heavy-quark operator matrix element defined in (A.6) and 
$ {\cal C}_{k,i}\,\, (k=2,3; i=q,g)$ denote the light-parton
coefficient functions which are presented
up to order $\alpha_s^2$ in Appendix B of \cite{zn}.
Here we want to emphasize that relations (\ref{eqn:42.17})-(\ref{eqn:42.19})
only hold in the asymptotic limit $Q^2 \gg m^2$.
Further they are presented in the $\overline {\rm MS}$-scheme.
One can check the validity of (\ref{eqn:42.17}) using the exact 
order $\alpha_s$ contributions to the coefficient functions
$H_{k,g}^{W}$ and $H_{2,g}^{\gamma}$ computed in \cite{got}.
Finally notice that the second order coefficient functions
${\cal C}_{k,i}^{(2)}$ require the knowledge of the three-loop DGLAP
splitting functions in order to make a complete next-to-next-to-leading
order (NNLO) analysis of $F_{k,c}^{\nu p}$ in (\ref{eqn:42.7}).
However these splitting functions are not known yet so that one cannot
obtain the parton densities in the next-to-next-to-leading log (NNLL)
approximation. Hence the order $\alpha_s^2$ contributions to the structure
functions $F_{k,c}^{\nu p}$ due to the expressions in
(\ref{eqn:42.18}), (\ref{eqn:42.19}) have to be considered
as an estimate of the exact QCD corrections beyond NLO.
However in practice it turns out that the contributions due to 
the coefficient functions are much more important than those 
coming from the higher order DGLAP splitting functions provided
both are represented in the $\overline {\rm MS}$-scheme (see \cite{timo}).

~

%% file: ch3.tex
\mysection{${\cal O}(\alpha_s^2)$ contributions to the 
charged-current process due to charm 
production}
\label{sec:ch3}

In this section we will discuss the QCD corrections to the charged-current
structure functions $F_{k,c}(x,Q^2,m^2)\,\, (k = 2,3)$ measured 
at fixed isoscalar target experiments :
$(\nu_{\mu} ({\bar \nu}_{\mu}) + N \rightarrow \mu^{-} (\mu^{+}) + '\!\!X')$
\cite{rab}-\cite{ccfr}
 and the collider experiments at HERA : $(e^-(e^+) + p \rightarrow
\nu_e ({\bar \nu}_e) + '\!\!X')$ \cite{ahm}-\cite{der}. The kinematical ranges
in which the fixed-target experiments are carried out are given by
$0.01 < x < 1$ and  $2 < Q^2 < 100~({\rm GeV}/c)^2$. For HERA they become 
$0.006 < x < 0.5$  and  $200 < Q^2 < 10^4\,({\rm GeV}/c)^2 $.
In particular we are interested in the order $\alpha_s^2$
contributions originating from the heavy-quark 
coefficient functions $L_{k,q}^{W,\rm NS, (2)}$ (exact) and
$H_{k,q}^{W,\rm PS, (2)}$, $H_{k,g}^{W,\rm S, (2)}$ (asymptotic)
presented in the previous section.
Since we are dealing with approximations to the 
above contributions only we have to determine the kinematical range in which
they are valid. First we have to investigate for which 
$x$- and $Q^2$-values
the order $\alpha_s$ corrections coming from 
$H_{k,g}^{W,\rm S, (1)}$ (W- boson-gluon fusion process)
dominate those which are due to 
$H_{k,q}^{W,\rm NS, (1)}$ (flavour-excitation process).
Then we have to determine $Q^2_{min}$ so that for $Q^2 > Q^2_{min}$
the results obtained from the asymptotic and exact expressions of 
the heavy-quark
coefficient function $H_{k,g}^{W,\rm S, (1)}$ coincide. Because
of the observations made in \cite{bmsn1}-\cite{bmsn2} for the one-photon
exchange process we expect that the same values of $Q^2_{min}$
also holds for the order $\alpha_s^2$ contributions
to the charged-current interactions given by
$H_{k,q}^{W,\rm PS, (2)}$, $H_{k,g}^{W,\rm S, (2)}$.

In our plots we will use the notations that $F_{k,c}^{(l)}$ $(l= 0,1,2)$
receive contributions from the coefficients corrected up to
order $\alpha_s^l$. This implies that for $F_{k,c}^{(0)}$ 
(Born-approximation or LO) we have to adopt the leading log (LL)
parton densities and the running coupling constant.
In the case of $F_{k,c}^{(1)}$ (NLO) we have to choose the next-to-leading
log (NLL) parton densities and the running coupling constant.
This will be also done for $F_{k,c}^{(2)}$ where we only have
included the order $\alpha_s^2$ contributions due to the 
heavy-quark coefficient functions $L_{k,q}^{W,\rm NS, (2)}$,
$H_{k,q}^{W,\rm PS, (2)}$ and $H_{k,g}^{W,\rm S, (2)}$.
All parton densities, coefficient functions and the running
coupling constant are presented in the $\overline{\rm MS}$-scheme.
Further we choose the factorization scale equal to the 
renormalization scale and set $\mu^2= Q^2$.
For other scales and the scale dependence of $F_{k,c}$ see \cite{kls},
\cite{bag}.
The structure functions in (\ref{eqn:42.7}) are presented in the 
fixed-flavour scheme (FFS) where the number of light flavours
is chosen to be $n_f = 3$. Since the parton density set GRV94 \cite{grv94}
is the only one which is consistently determined in the 
FFS we will adopt this in our paper. In LO and NLO the QCD scale 
$\Lambda_{n_f}$, which has to be substituted in the running coupling
constant, is given by $\Lambda_3^{(\rm LO)} = 248~{\rm MeV}$ and
$\Lambda_3^{(\rm NLO)} = 232~{\rm MeV}$ respectively.
The remaining input parameters are the charm-quark mass and the 
Cabibbo angle which are chosen to be $m = 1.5 \,({\rm GeV}/c)$ 
and $| \sin \theta_C| = 0.221$
\cite{barn}. For the dependence of $F_{k,c}$ on $m$ see \cite{kl}.

In order to determine the kinematical range in which the 
approximations are valid we will first plot the quantities 
$F_{k,c}^{q}$ and $F_{k,c}^{g}$ which receive the order $\alpha_s$
contributions from $H_{k,q}^{W,\rm NS, (1)}$ and
$H_{k,g}^{W,\rm S, (1)}$ respectively. 
In Fig.~1 we have presented $F_{2,c}^q$, $F_{2,c}^g$ for an
isoscalar target in neutrino reactions at $Q^2 = 10,100 \,({\rm GeV}/c)^2$. 
Here one observes that the W-boson-gluon fusion process 
$(F_{2,c}^g)$ dominates the flavour-excitation process $(F_{2,c}^q)$
when $x < 0.07$ $(Q^2 = 10\,({\rm GeV}/c)^2)$. At larger $Q^2$ i.e. 
$(Q^2 = 100~({\rm GeV}/c)^2)$ $x$ increases and the latter becomes equal to
$x=0.2$.
The same plots are given for $F_{3,c}^q$ and $F_{3,c}^g$ in Fig.~2.
It turns out that for $F_{3,c}$ the $x$-values, below which 
$F_{3,c}^g$ dominates $F_{3,c}^q$, are larger that the ones
obtained for $F_{2,c}$. In Fig.~2 they 
become $x < 0.3$  for both $Q^2 = 10~({\rm GeV}/c)^2$ and 
$Q^2 = 100~({\rm GeV}/c)^2$.
{}From the above we conclude that already at rather low $Q^2$-values 
the W-boson-gluon fusion mechanism completely dominates the order $\alpha_s$
corrections to $ F_{2,c}$ provided $x < 0.1$.
This also holds for $ F_{3,c}$ at even larger $x$-values i.e. $x < 0.3$.

Now we have to determine at  which value of $Q^2$, denoted by $Q^2_{min}$,
the exact heavy-quark coefficient function $H_{k,g}^{W,\rm S,(1)}$
can be replaced by its asymptotic expression without
altering the structure functions $F_{k,c}$.
For that purpose we will study the quantity

\begin{eqnarray}
\label{eqn:43.1}
R_{k,c}(x,Q^2,m^2) = \frac{F_{k,c}^{g,asymp}(x,Q^2,m^2)}
                           {F_{k,c}^{g,exact}(x,Q^2,m^2)}
\end{eqnarray}
where the numerator and the denominator are computed
using the asymptotic
and the exact expressions for $H_{k,g}^{W,\rm S, (1)}$ respectively.
Further we focus our attention on the range $0.006 < x < 0.1$ which is
explored by the fixed-target experiments and the HERA collider. The range
$0.1 \ge x<1$ falls beyond our scope because here the $W$-boson-gluon fusion
process is not dominant anymore.
In Fig.~3 we have plotted $R_{2,c}$ 
in the range $10 < Q^2 < 10^4~({\rm GeV}/c)^2$ 
for $x = 0.1,0.01,10^{-3}, 10^{-4}$.
The same is done for $R_{3,c}$ in Fig.~4. Fig.~3 reveals that for
$x= 0.1$, $R_{2,c}> 0.9$ when $Q^2 > Q^2_{min}= 50~({\rm GeV}/c)^2$.
At decreasing $x$ the value of $Q^2_{min}$ will become less so that
for $x= 10^{-4}$ one obtains $Q^2_{min}= 20~({\rm GeV}/c)^2$. 
A similar behaviour is shown by Fig.~4. Here we observe that for $x=0.1$,
$R_{3,c} > 0.9$ when $Q^2_{min}=85~({\rm GeV}/c)^2$. Also in this case
$Q^2_{min}$ decreases if $x$ gets lower. For instance for
$x = 10^{-4}$ we get $Q^2_{min}=50~({\rm GeV}/c)^2$ .
{}From the above one cannot conclude that the replacement of the exact
by the asymptotic coefficient functions works better for
$F_{2,c}$ than for $F_{3,c}$. This becomes clear when Fig.~5 
is compared with Fig.~6 where now $R_{k,c}$ (\ref{eqn:43.1}) 
$(k = 2,3)$ is plotted
as a function of $x$ for $Q^2 = 10,20,50,100~({\rm GeV}/c)^2$. If
we choose $Q^2 = 10~({\rm GeV}/c)^2$ then in the range $0.01 < x < 0.1$ we
observe that $0.6 > R_{3,c} > 0.5$ (Fig.~6). 
However the behaviour of $R_{2,c}$ is terrible. In the 
range $0.01 < x < 0.05$ we find $0.7 > R_{2,c}> 0.1$. 
For $x > 0.05$, $R_{2,c}$ becomes even negative and for 
very large $x$ it switches sign again. This behaviour of the coefficient
functions originates from mass factorization of the original partonic
cross sections which destroyes the positive definiteness of the latter
quantities.
Further we also studied the behaviour of 
$F_{k,c}^q$, $F_{k,c}^g$ and $R_{k,c}$ for antineutrino isoscalar
reactions but the results are the same as found above. 
Summarizing our findings we can state that for $x < 0.1$ and
$Q^2 > Q^2_{min} = 50~({\rm GeV}/c)^2$ the
approximation to the structure functions, given by the replacement of the exact 
by the asymptotic heavy-quark coefficient functions, works quite well.
One can also choose smaller values for $Q^2_{min}$. However in this case
the approximation will still work for $F_{3,c}$ but it becomes bad for 
$F_{2,c}$ unless one chooses smaller values for $x$. For example for
$Q^2_{min} = 20~({\rm GeV}/c)^2$ one has to take $x < 0.08$.
{} From now on we will assume that the same $Q^2_{min}$ and $x$
also hold for the order $\alpha_s^2$ contributions
to the heavy-quark coefficient functions $H_{k,q}^{W,\rm PS}$, 
$H_{k,g}^{W,\rm S}$
This assumption is based on our findings in \cite{bmsn1}-\cite{bmsn2}
for the electromagnetic-current process where the above
contributions are exactly known up to order $\alpha_s^2$.

In the next figures we show $F_{2,c}^{(l)}$ $(l = 0,1,2)$ for the
neutrino isoscalar target reaction. Here
we have plotted $F_{2,c}^{(0)}$(exact),  $F_{2,c}^{(1)}$(exact) 
and  $F_{2,c}^{(2)}$
(approximate) for $Q^2 = 10~({\rm GeV}/c)^2$ (Fig.~7)
and $Q^2= 100~({\rm GeV}/c)^2$ (Fig.~8).
Fig.~7 reveals that the order $\alpha_s^2$ contributions, which are
positive, become very small. However here one has to bear in mind
that for $x \ge 0.03$ and $Q^2_{min} = 10~({\rm GeV}/c)^2$
the estimate of this contribution is not very reliable
in view of what we found  in Fig.~5. 
At $Q^2 = 100~({\rm GeV}/c)^2$ (Fig.~8), 
where the approximation is very accurate,
the order $\alpha_s^2$ corrections are larger and at $x=0.01$ they
 amount to $5 \%$
of $F_{2,c}^{(1)}$ which is still quite small. These
corrections in the case of $F_{3,c}$ (see Figs.~9,10) 
become much larger  than the ones observed for $F_{2,c}$.
Furthermore in the light of the discussion above for 
$R_{3,c}$ (Fig.~6) they are more reliable even for
 $Q^2 = 10~({\rm GeV}/c)^2$
(Fig.~9). Like in the case of $F_{2,c}$ the largest
corrections occur at the lowest $x$-value and they decrease when $x$ tends
to one. At $x = 0.01$ the order $\alpha_s^2$ corrections w.r.t.
$F_{3,c}^{(1)}$ amount to $14\%$ for $Q^2 = 10~({\rm GeV}/c)^2$ (Fig.~9)
and $20\%$ for $Q^2 = 100~({\rm GeV}/c)^2$ (Fig.~10).
Therefore at increasing $Q^2$ the corrections get larger.
Finally we want to call attention to the difference in 
behaviour between $F_{2,c}^{(l)}$ and  $F_{3,c}^{(l)}$
for $l= 0,1,2$. From Figs.~7-10 we infer that
$F_{2,c}^{(2)} > F_{2,c}^{(1)} > F_{2,c}^{(0)}$
whereas $F_{3,c}^{(0)} > F_{3,c}^{(2)} > F_{3,c}^{(1)}$.

We will now investigate how accurately one can determine the
strange-quark density $s(x,\mu^2)$ from the data
obtained for the charged-current \cite{olt}
and electromagnetic-current \cite{arn}
interactions.
To that order we will compute the ratio in the three-flavour
scheme which is defined by

\begin{eqnarray}
\label{eqn:43.2}
R_{s}^{(l)}(x, Q^2, m^2) = \frac{\Delta^{(l)}(x, Q^2, m^2)}
                                { x s^{(l)}(x, Q^2)}
\end{eqnarray}
with (see also (\ref{eqn:41.6}))

\begin{eqnarray}
\label{eqn:43.3}
\Delta^{(l)}& = & \frac{5}{12} \Biggl(
               F_2^{\nu N, (l)} +  F_2^{\bar {\nu} N, (l)}
               +  F_{2,c}^{\nu N, (l)} +  F_{2,c}^{\bar {\nu} N, (l)}
               \Biggr)
\nonumber \\ &&
               - 3 \Biggl( 
               F_2^{\mu N, (l)} + F_{2,c}^{\mu N, (l)} \Biggr)
\end{eqnarray}
where $F_{2}^{l N}$, $F_{2,c}^{l N}$ $(l= \nu, \bar{\nu}, \mu)$
denote the light-parton contributions ((\ref{eqn:42.1}), (\ref{eqn:42.14}))
and the charm component ((\ref{eqn:42.7}), (\ref{eqn:42.15}))
of the structure functions respectively.
For the computation of $F_{2,c}^{\mu N}$ we will use the exact
order $\alpha_s^2$ contributions to the heavy-quark coefficient
functions as presented in \cite{lrsn}.
Further $s^{(0)}(x, \mu^2)$ and  $s^{(1)}(x, \mu^2)$ 
stand for the LL and NLL parametrizations of the strange-quark
density. Since the NNLL parametrization is unknown we have 
to put $s^{(2)}(x, \mu^2) = s^{(1)}(x, \mu^2)$.
The phenomenon that $R_{s}^{(0)}\not 0$ can be
 only due to the effect of the mass
of the charm quark. For $l \geq 1$ one gets in addition
deviations which originate from the higher order QCD
corrections. We have plotted $R_{s}^{(l)} $ $(l = 0,1,2)$ in the range
$10^{-4} < x< 1$ for three different $Q^2$-values i.e.
$Q^2 = 10,100,10^4~({\rm GeV}/c)^2$.
In Fig.~11 $(Q^2 = 10~({\rm GeV}/c)^2)$ 
we observe that $R_s^{(0)}$ strongly deviates from one
in the region $0.1< x <1$, which is wholly due to 
thresholds effects of the charm quark in $F_{2,c}^{\nu N,(0)}$ and
$F_{2,c}^{\bar{\nu} N,(0)}$. 
This effect will vanish when $Q^2$ gets larger (see Figs.~12,13).
Notice that in the region $x \ge 0.1$ the size of the strange-quark
density is very small. For $x < 0.1$ $R_{s}^{(0)}$ tends to one 
for $Q^2> 10~({\rm GeV}/c)^2$.  
This picture does not change when one includes
the order $\alpha_s$ corrections provided $R_s^{(1)}$ is 
computed at small $Q^2$ (see Fig.~11). At large $Q^2$ (see Figs.~12,13),
$R_{s}^{(1)}$ differs apprecibly from one in particular at small $x$.
The difference becomes even larger when one includes the
order $\alpha_s^2$ contributions (see $R_s^{(2)}$ in Figs.~11-13).
This effect is wholly due to the heavy-quark coefficient functions
$H_{2,i}^{V} ~ (V = \gamma, W)$, appearing in 
the structure functions of (\ref{eqn:43.3})
which grow as $\ln^i(Q^2/m^2)$ when $Q^2\gg m^2$.
Notice that from Fig.~5 one can see that the estimate of the
order $\alpha_s^2$ contributions to 
$F_{2,c}^{\nu N}$ and $F_{2,c}^{\bar{\nu} N}$  is quite reliable
for $Q^2 > 10~({\rm GeV}/c)^2$ and $x < 0.01$. 
{}From Figs.~11-13 we can conclude that at small $x$, $R_s^{(1)}$
and $R_s^{(2)}$ become smaller than 0.5, which means that the actual 
strange-quark density is more than two times larger than the value 
obtained for $\Delta$ in (\ref{eqn:43.3}). Hence one can conclude
that the higher order QCD corrections bedevil the extraction
of the strange-quark density in the region $x<0.1$.
Furthermore the effect of the mass of the charm quark is not
negligible at small $Q^2$ $(Q^2 < 10~({\rm GeV}/c)^2)$ 
in the  large $x$-region
$(x \ge 0.1)$.

The steep rise of the heavy-quark coefficient functions
which grow like $\ln^i(Q^2/m^2)$  for $Q^2\gg m^2$
becomes even more conspicuous if we study 
charm production at HERA where one measures the structure
functions $F_{k,c}^{\bar{\nu_e} p} (\equiv F_{k,c}^{e^- p})$ and
$F_{k,c}^{\nu_e p} ( \equiv F_{k,c}^{e^+ p})$.
The latter are plotted for 
$Q^2= 10^4~({\rm GeV}/c)^2$  in the
region $0.006 < x < 0.5$ in Figs.~14-17.
In this kinematical region the approximations to the second order
heavy-quark coefficient functions
become excellent (see Figs.~3-6).
This already happens at the lowest $Q^2$-value i.e. 
$Q^2 = 200~({\rm GeV}/c)^2$
measured in \cite{der}.
In the case of $F_{2,c}$ the exact order $\alpha_s^2$ corrections 
are not that spectacular and they amount to almost $7\%$ at
$x = 0.006$ w.r.t. to $F_{2,c}^{(1)}$ (see Figs.~14,15).
This is in contrast to $F_{3,c}$ (see Figs.~16,17) where
the order $\alpha_s^2$ corrections become $32\%$ w.r.t.
$F_{3,c}^{(1)}$ (see Fig.~17).
These large corrections, occurring at small
$x$ and large $Q^2$, vitiate the perturbation series in particular
the one given for $F_{3,c}$.
Therefore the large logarithmic terms represented by
$\ln^i(Q^2/m^2)$  have to be resummed according to the VFNS
approach (see \cite{acot}, \cite{bmsn2}). 
In the latter scheme the charm component of the
structure functions takes the following form 

\begin{eqnarray}
\label{eqn:43.4}
&&  \hskip -0.8cm
F_{k,c}^{W^+ p} (x, Q^2) \equiv F_{k,c}^{\nu p} (x, Q^2)   =
\nonumber \\ && \hskip -0.8cm
  a_k(x) \int_x^1 \frac{d \, z}{z}
\Biggl[\Sigma_4\Biggl(\frac{x}{z}, \mu^2\Biggr) 
\tilde {\cal C}_{k,q}^{\rm PS} \Biggl(z, \frac{Q^2}{\mu^2}\Biggr)
+
G\Biggl(\frac{x}{z}, \mu^2\Biggr) 
\tilde {\cal C}_{k,g}^{\rm S} \Biggl(z, \frac{Q^2}{\mu^2}\Biggr) +
\nonumber \\ && \hskip -0.8cm
\Biggl\{ b_{k,\bar{c}} \bar{c} \Biggl(\frac{x}{z}, \mu^2\Biggr)
+ 
b_{k,d} \sin^2(\theta_C) d \Biggl(\frac{x}{z}, \mu^2\Biggr)
+b_{k,s} \cos^2(\theta_C) s \Biggl(\frac{x}{z}, \mu^2\Biggr)
\Biggr\} 
{\cal C}_{k,q}^{\rm NS} \Biggl(z, \frac{Q^2}{\mu^2}\Biggr)\Biggr]
\nonumber \\ && \hskip -0.8cm
\,
\end{eqnarray}
where the parton densities and the light-quark coefficient functions 
${\cal C}_{k,i}$ $(k = 2,3; i= q,g)$ are presented in the
four-flavour scheme.
The parton-density set now also contains the
charm-quark density $c(z,\mu^2)$ and the
singlet combination reads

\begin{eqnarray}
\label{eqn:43.5}
\Sigma_4(z,\mu^2) = \Sigma_3(z,\mu^2) + c(z,\mu^2) + \bar{c}(z,\mu^2) 
\end{eqnarray}
where $\Sigma_3(z,\mu^2)$ is defined in (\ref{eqn:42.4}).
The structure function $F_{k,c}^{\bar{\nu} p}$ ($\equiv F_{k,c}^{W^- p}$)
can be derived by 
interchanging quarks and anti-quarks (see below (\ref{eqn:42.9})).
In order to compute (\ref{eqn:43.4}) we need the charm-quark
density which however is not present in GRV94 \cite{grv94}
because the latter set is presented in a three-flavour scheme.
Therefore we have adopted GRV92 \cite{grv92} with $\Lambda_4^{(\rm LO)}= 
\Lambda_4^{(\rm NLO)} = 200~{\rm MeV}$
which contains the charm-quark density.
We first checked that using the densities in GRV92 for the
expressions in the three-flavour scheme one gets the
same results for $F_{2,c}$ as obtained with GRV94.
In the case of $|F_{3,c}|$, the latter set produces
values which are $10\%$ below those given by GRV92. As a comparison
we have also plotted in Figs.~14-17 $F_{k,c}^{(1)}$ (VFNS)  
using Eq.~(\ref{eqn:43.4}) and
GRV92. It turns out that the result for  
$F_{2,c}^{(1)}$ is larger in VFNS (\ref{eqn:43.4}) than obtained by FFS
(see Figs.~14,15). This observation was also made in \cite{bag}. 
The discrepancy becomes less when the order $\alpha_s^2$ corrections are
included in the case of FFS (see $F_{2,c}^{(2)}$).
This feature was also discovered for $F_{2,c}^{e p}$ in the
electromagnetic-current process \cite{bmsn2}.
However, for $F_{3,c}$ the difference between the two schemes
is large (see Figs.~16,17), in particular
in the small $x$-region. Here we observe that 
$F_{3,c}^{(1)}$ in VFNS deviates
appreciably from $F_{3,c}^{(1)}$ as well as $F_{3,c}^{(2)}$ in FFS.
This is no surprise because, as we have seen above, in FFS the large logarithms
in the heavy-quark coefficient functions lead to larger corrections to
$F_{3,c}$ than to $F_{2,c}$.
Therefore the resummation of these terms will have a larger effect
on $F_{3,c}$ than on $F_{2,c}$.
This is the explanation of the difference between the 
predictions of the FFS and VFNS approach for the
HERA experiments as originally observed for $F_{2,c}^{(1)}$ in \cite{bag}.
Since we believe that the resummation of these 
large logarithmic terms has to be carried out for 
$Q^2 \gg m^2$, the results obtained by Eq.~(\ref{eqn:43.4})
(VFNS) are much more reliable than those obtained in FFS
given by the expressions in Section~\ref{sec:ch2}.

Summarizing our findings, we have studied the effect of the 
order $\alpha_s^2$ contributions from the heavy-quark
coefficient functions to the charm component of the
charged-current structure functions.
Since for some of them the exact expressions do not exist we have
to rely on approximations which are strictly
speaking valid only for $Q^2 \gg m^2$ (HERA-regime).
However, from a comparison based on the exact and asymptotic order
$\alpha_s$ heavy-quark coefficient functions
corresponding to the W-boson-gluon fusion process,
the predictions are also reliable for 
$Q^2 > 20~({\rm GeV}/c)^2$ and $x < 0.08$.
At smaller $x$ it is even sufficient to choose $Q^2 > 10~({\rm GeV}/c)^2$.

The outcome of our analysis reveals that the order $\alpha_s^2$
corrections are small for $F_{2,c}$ but large for $F_{3,c}$, in
particular when $Q^2$ is big. These large corrections are due
to the logarithmic terms $\ln^i(Q^2/m^2)$ dominating the 
heavy-quark coefficient functions characteristic of the FFS approach.
Therefore at large $Q^2$ these terms have to be resummed in the
context of VFNS, leading to a better prediction, in particular
for $F_{3,c}$.
Finally we want to stress that the higher order QCD
corrections bedevil the extraction of the strange-quark
density from the charged- and electromagnetic-current  
total structure functions, in particular at small $x$ $(x<0.08)$.


%% file: cha.tex
\mysection*{Appendix A}
\setcounter{section}{1}

In this section we will derive the relations between the
heavy-quark coefficient functions $H_{k,i}^{W}$ and $H_{2,i}^{\gamma}$
 $(k= 2,3; i= q,g)$ obtained for the charged-current and the 
electromagnetic-current processes respectively.
Using mass-factorization techniques one has shown in \cite{bmsmn}-\cite{bmsn2}
that there exists a relation between the asymptotic $(Q^2\gg m^2)$
heavy-quark coefficient functions $H_{k,i}^{V}$ and the light-parton
coefficient functions ${\cal C}_{k,i}$. 
Hence one can derive  that up to order $\alpha_s^2$  the heavy-quark
coefficient functions corresponding to the reactions

\begin{eqnarray}
\label{eqn:4A.1}
i + V \rightarrow q_1 + \bar{q}_2
\, ,
\end{eqnarray}
\begin{eqnarray}
\label{eqn:4A.2}
i + V \rightarrow q_1 + \bar{q}_2 + j
\, ,
\end{eqnarray}
in the asymptotic limit $Q^2 \gg m^2$ can be written as (see Eqs.~ (2.34),
(2.35) in \cite{bmsn2}) 

\begin{eqnarray}
\label{eqn:4A.3}
&& H_{k,i}^{V}\Biggl( \frac{Q^2}{m_1^2}, \frac{Q^2}{m_2^2},
                   \frac{m_1^2}{\mu^2}, \frac{m_2^2}{\mu^2} \Biggr)
 = 
\frac{1}{2} A_{q_1 i}\Biggl( \frac{m_1^2}{\mu^2}\Biggr)
\otimes {\cal C}_{k,q_1}^{\rm S} \Biggl( \frac{Q^2}{\mu^2}  \Biggr)
\nonumber \\ &&
+\frac{1}{2} A_{\bar{q_2} i} \Biggl( \frac{m_2^2}{\mu^2}\Biggr)
\otimes {\cal C}_{k,\bar{q_2}}^{\rm S} \Biggl( \frac{Q^2}{\mu^2}  \Biggr)
+\frac{1}{2} A_{g i}\Biggl( \frac{m_1^2}{\mu^2}\Biggr)
\otimes \tilde {\cal C}_{k,g}^{\rm S}\Biggl( \frac{Q^2}{\mu^2}  \Biggr)
\nonumber \\ &&
+\frac{1}{2} A_{g i}\Biggl( \frac{m_2^2}{\mu^2}\Biggr)
\otimes \tilde {\cal C}_{k,g}^{\rm S}\Biggl( \frac{Q^2}{\mu^2}  \Biggr)
\, .
\end{eqnarray}
For convenience we have suppressed the Bjorken scaling variable 
$z=Q^2/2 p_i q$ and $\otimes$ denotes the convolution symbol defined by
\begin{eqnarray}
\label{eqn:4A.4}
\Big(f\otimes g\Big)(z)=\int_0^1 dz_1\int_0^1 dz_2  ~
\delta(z-z_1z_2)f(z_1)g(z_2)  ~.
\end{eqnarray}

The light partons involved in reactions (\ref{eqn:4A.1}) and (\ref{eqn:4A.2})
are given by $i$ and $j$ $(i,j=q,g)$. For $V=\gamma$ both
$q_1$ and $q_2$ stand for the heavy quarks whereas for $V=W$ one of them
is represented by a light quark.
The masses of $q_1$ and $q_2$ are given by $m_1$ and $m_2$ respectively.
The light-parton coefficient functions ${\cal C}_{k,i}$ $(i=q,g)$ are
obtained from the massless-parton subprocesses
\begin{eqnarray}
\label{eqn:4A.5}
i + V \rightarrow j_1 + j_2 +\dots + j_k
\end{eqnarray}
which are independent of the nature of the probe $V$. Further when the
quarks $q_1$ and $q_2$ in (\ref{eqn:4A.3}) are massless the coefficient 
functions satisfy the relation ${\cal C}_{k,q_1}^{\rm S}
={\cal C}_{k,q_2}^{\rm S}$. The light parton coefficient functions in
(\ref{eqn:4A.3}) are calculated up to order $\alpha_s^2$ in \cite{zn}.
Expression (\ref{eqn:4A.3}) requires the knowledge of the 
renormalized heavy-quark operator matrix elements (OME's)
\begin{eqnarray}
\label{eqn:4A.6}
A_{q_l i} \Biggl(\frac{m_l^2}{\mu^2}\Biggr) = <i| O_{q_l} (0)|i>
\end{eqnarray}
and the heavy-quark loop contributions to the renormalized light-parton 
OME's given by
\begin{eqnarray}
\label{eqn:4A.7}
A_{j i} \Biggl(\frac{m_l^2}{\mu^2}\Biggr) = <i| O_{j} (0)|i>
\end{eqnarray}
with 
\begin{eqnarray}
\label{eqn:4A.8}
A_{j i}^{(0)} \Biggl(z, \frac{m_l^2}{\mu^2}\Biggr) = 
\delta_{j i} \delta(1 - z)
\, .
\end{eqnarray}
The above OME's are calculated up to order $\alpha_s^2$ in \cite{bmsmn}
and the renormalized expressions are presented in the $\overline{\rm MS}$-scheme
in Appendix B of \cite{bmsn2}.
Furthermore they satisfy the conditions when $m_l^2=0$
\begin{eqnarray}
\label{eqn:4A.9}
A_{q_l i}(0) = 0 , \quad \quad A_{j i}(z,0) = \delta_{j i} \delta(1 - z)
\, .
\end{eqnarray}
The local operators $O_j(x)$ appear in the light-cone expansion
of the product of the two electroweak currents which, after having
taken the Fourier transform, show up in the expressions for the
structure functions $F_k(x,Q^2)$ (see \cite{pol}).

In the case of $V=\gamma$ Eq.~(\ref{eqn:4A.3}) leads in the first and second
order to the following relations $(m_1=m_2=m; q_1=q_2=c)$
\begin{eqnarray}
\label{eqn:4A.10}
H_{2,g}^{\gamma,(1)} \Biggl( \frac{Q^2}{m^2}, \frac{m^2}{\mu^2} \Biggr) = 
A_{c g}^{(1)}  \Biggl(\frac{m^2}{\mu^2} \Biggr)
+
\tilde {\cal C}_{2,g}^{\rm S,(1)}  \Biggl(\frac{Q^2}{\mu^2} \Biggr)
\, ,
\end{eqnarray}

\begin{eqnarray}
\label{eqn:4A.11}
H_{2,g}^{\gamma,(2)} \Biggl( \frac{Q^2}{m^2}, \frac{m^2}{\mu^2} \Biggr)  & = &
A_{c g}^{(2)}  \Biggl(\frac{m^2}{\mu^2} \Biggr)
+ A_{c g}^{(1)}  \Biggl(\frac{m^2}{\mu^2} \Biggr)
\otimes {\cal C}_{2,q}^{\rm NS,(1)}  \Biggl(\frac{Q^2}{\mu^2} \Biggr)
\nonumber  \\ && 
+ \tilde {\cal C}_{2,g}^{\rm S,(2)}  \Biggl(\frac{Q^2}{\mu^2} \Biggr)
\, ,
\end{eqnarray}

\begin{eqnarray}
\label{eqn:4A.12}
H_{2,q}^{\gamma,\rm PS,(2)} \Biggl( \frac{Q^2}{m^2}, \frac{m^2}{\mu^2} \Biggr)
=
A_{c q}^{\rm PS,(2)} \Biggl(\frac{m^2}{\mu^2} \Biggr)
+
\tilde {\cal C}_{2,q}^{\rm PS,(2)} \Biggl(\frac{Q^2}{\mu^2} \Biggr)
\, ,
\end{eqnarray}
where we have expanded all quantities up to order $(\alpha_s^2/4\pi)^2$.
Notice that in the case of $V=\gamma$, $H_{3,i}^{\gamma}=0$.
Furthermore ${\cal C}_{i,c}={\cal C}_{i,q}$ when $m \rightarrow 0$.

For $V=W$ the above relations become $(m_1=m,m_2=0; q_1=c,~q_2=d,s)$

\begin{eqnarray}
\label{eqn:4A.13}
H_{2,g}^{W,(1)} \Biggl( \frac{Q^2}{m^2}, \frac{m^2}{\mu^2} \Biggr) =
\frac{1}{2} A_{c g}^{(1)}  \Biggl(\frac{m^2}{\mu^2} \Biggr)
+
\tilde {\cal C}_{2,g}^{\rm S,(1)}  \Biggl(\frac{Q^2}{\mu^2} \Biggr)
\, ,
\end{eqnarray}

\begin{eqnarray}
\label{eqn:4A.14}
H_{2,g}^{W,(2)} \Biggl( \frac{Q^2}{m^2}, \frac{m^2}{\mu^2} \Biggr) & = &
\frac{1}{2} A_{c g}^{(2)}  \Biggl(\frac{m^2}{\mu^2} \Biggr)
+ \frac{1}{2} A_{c g}^{(1)}  \Biggl(\frac{m^2}{\mu^2} \Biggr)
\otimes {\cal C}_{2,q}^{\rm NS,(1)}  \Biggl(\frac{Q^2}{\mu^2} \Biggr)
\nonumber \\ &&
+ \tilde {\cal C}_{2,g}^{\rm S,(2)}  \Biggl(\frac{Q^2}{\mu^2} \Biggr)
\, ,
\end{eqnarray}

\begin{eqnarray}
\label{eqn:4A.15}
H_{2,q}^{W,\rm PS,(2)} \Biggl( \frac{Q^2}{m^2}, \frac{m^2}{\mu^2} \Biggr)
=
\frac{1}{2} A_{c q}^{\rm PS,(2)} \Biggl(\frac{m^2}{\mu^2} \Biggr)
+
\tilde {\cal C}_{2,q}^{\rm PS,(2)} \Biggl(\frac{Q^2}{\mu^2} \Biggr)
\, .
\end{eqnarray}
For the heavy-quark coefficient functions $H_{3,i}^{W}$ we have to use the 
following relations

\begin{eqnarray}
\label{eqn:4A.16}
{\cal C}_{3,q}^{\rm NS} = - {\cal C}_{3,\bar{q}}^{\rm NS} , \quad
\tilde {\cal C}_{3,g}^{\rm S} =0, \quad
\tilde {\cal C}_{3,q}^{\rm PS} =0
\end{eqnarray}
which follow from charge conjugation invariance of the
strong interactions.
Hence we can derive from Eq.~(\ref{eqn:4A.3})

\begin{eqnarray}
\label{eqn:4A.17}
H_{3,g}^{W,(1)} \Biggl( \frac{Q^2}{m^2}, \frac{m^2}{\mu^2} \Biggr)
=
\frac{1}{2} A_{c g}^{(1)} \Biggl(\frac{m^2}{\mu^2} \Biggr)
\, ,
\end{eqnarray}

\begin{eqnarray}
\label{eqn:4A.18}
&& \hskip -1.0cm 
H_{3,g}^{W,(2)} \Biggl( \frac{Q^2}{m^2}, \frac{m^2}{\mu^2} \Biggr)  = 
\frac{1}{2} A_{c g}^{(2)}  \Biggl(\frac{m^2}{\mu^2} \Biggr)
%
+ \frac{1}{2} A_{c g}^{(1)}  \Biggl(\frac{m^2}{\mu^2} \Biggr)
\otimes {\cal C}_{3,q}^{\rm NS,(1)}  \Biggl(\frac{Q^2}{\mu^2} \Biggr)
\, ,
\end{eqnarray}

\begin{eqnarray}
\label{eqn:4A.19}
H_{3,q}^{W,\rm PS,(2)} \Biggl( \frac{Q^2}{m^2}, \frac{m^2}{\mu^2} \Biggr)
=
\frac{1}{2} A_{c q}^{\rm PS,(2)} \Biggl(\frac{m^2}{\mu^2} \Biggr)
\, .
\end{eqnarray}
Similar relations can be derived when $m_1=0, m_2=m$ and $ q_1=d,s;~q_2=c$.
Using Eqs.~(\ref{eqn:4A.10})-(\ref{eqn:4A.12})
one can now express $H_{k,i}^{W}$ into $H_{k,i}^{\gamma}$.
The results are presented in (\ref{eqn:42.17})-(\ref{eqn:42.19}).


%% file: chr.tex
%

%% file: chf.tex
\centerline{\bf \large{Figure Captions}}

\begin{description}
\item[Fig. 1.]
The order $\alpha_s$ contributions to $F_{2,c}^{\nu N}$ (\ref{eqn:42.7})
due to flavour excitation : $F_{2,c}^q$ (\ref{eqn:41.4}) and $W$-boson-gluon
fusion : $F_{2,c}^g$ (\ref{eqn:41.5});\\
solid line: $F_{2,c}^q$ at $Q^2 = 10$ (GeV/$c)^2$, 
dotted line: $F_{2,c}^q$ at $Q^2 = 100$ (GeV/$c)^2$, 
dashed line: $F_{2,c}^g$ at $Q^2 = 10$ (GeV/$c)^2$, 
dashed-dotted line: $F_{2,c}^g$ at $Q^2 = 100$ (GeV/$c)^2$. 
\\
\item[Fig. 2.]
Same as Fig. 1 but now for $F_{3,c}^{\nu N}$ (\ref{eqn:42.7}) plotted in
absolute value.
\\
\item[Fig. 3.]
$R_{2,c}$ (\ref{eqn:43.1}) plotted as a function of $Q^2$ at fixed $x$;\\
$x = 10^{-1}$  (dotted line), $x = 10^{-2}$ (dashed-dotted line),
$x = 10^{-3}$ (dashed line) and $x = 10^{-4}$ (solid line).
\\
\item[Fig. 4.]
Same as Fig. 3 but now for $R_{3,c}$ (\ref{eqn:43.1}).
\\
\item[Fig. 5.]
$R_{2,c}$ (\ref{eqn:43.1})
plotted as a function of $x$ at fixed $Q^2$;\\
$Q^2 = 10$ (GeV/$c)^2$ (dotted line), $Q^2 = 20$ (GeV/$c)^2$
(dashed-dotted line), $Q^2 = 50$ (GeV/$c)^2$ (dashed line),
 $Q^2 = 100$ (GeV/$c)^2$ (solid line).
\\
\item[Fig. 6.]
Same as Fig. 5 but now for $R_{3,c}$ (\ref{eqn:43.1}).
\\
\item[Fig. 7.]
The order $\alpha_s^l$ corrected 
structure function  $F_{2,c}^{\nu N}$ (\ref{eqn:42.7}),
denoted by $F_{2,c}^{(l)}$, 
as a function of $x$ at $Q^2 =10 $ (GeV/$c)^2$;\\
dotted line : $F_{2,c}^{(0)}$,
dashed line : $F_{2,c}^{(1)}$,
solid line : $F_{2,c}^{(2)}$.
\\
\item[Fig. 8.]
Same as Fig. 7 but now at $Q^2 =100$  (GeV/$c)^2$.
\item[Fig. 9.]
Same as Fig. 7 but now for $F_{3,c}^{\nu N}$ (\ref{eqn:42.7})
at $Q^2 =10$  (GeV/$c)^2$.
\item[Fig. 10.]
Same as Fig. 7 but now for $F_{3,c}^{\nu N}$ (\ref{eqn:42.7})
at $Q^2 =100$  (GeV/$c)^2$.
\item[Fig. 11.]
The order $\alpha_s^l$ corrected
ratio  $R_s$ (\ref{eqn:43.2}),
denoted by $R_s^{(l)}$,
as a function of $x$ at $Q^2 =10 $ (GeV/$c)^2$;\\
solid line : $R_s^{(0)}$,
dashed line : $R_s^{(1)}$,
dotted line : $R_s^{(2)}$.
\\
\item[Fig. 12]
Same as Fig. 11 but now at $Q^2 =100 $ (GeV/$c)^2$.
\item[Fig. 13]
Same as Fig. 11 but now at $Q^2 =10^4 $ (GeV/$c)^2$.
\item[Fig. 14]
The order $\alpha_s^l$ corrected structure function
$F_{2,c}^{e^- p}$ ( $\equiv F_{2,c}^{\bar{\nu}_e p}$) (\ref{eqn:42.7}),
denoted by $F_{2,c}^{(l)}$,
as a function of $x$ at $Q^2 =10^4 $ (GeV/$c)^2$;\\
dotted line : $F_{2,c}^{(0)}$ (FFS),
dashed line : $F_{2,c}^{(1)}$ (FFS),
solid line : $F_{2,c}^{(2)}$ (FFS). As a comparison we have also shown
dashed-dotted line : $F_{2,c}^{(1)}$ (VFNS).
\\
\item[Fig. 15]
Same as Fig. 14 but now for $F_{2,c}^{e^+ p}$ 
( $\equiv F_{2,c}^{\nu_e p}$) (\ref{eqn:42.7}).
\\
\item[Fig. 16]
Same as Fig. 14 but now for $F_{3,c}^{e^- p}$
( $\equiv F_{2,c}^{\bar{\nu}_e p}$) (\ref{eqn:42.7}).
\\
\item[Fig. 17]
Same as Fig. 14 but now for $F_{3,c}^{e^+ p}$
( $\equiv F_{2,c}^{\nu_e p}$) (\ref{eqn:42.7}).

\end{description}